\documentclass[fleqn,usenatbib]{mnras}

\usepackage{newtxtext,newtxmath}

\usepackage[T1]{fontenc}
\usepackage{ae,aecompl}

\usepackage{graphicx}	
\usepackage{amsmath}	
\usepackage{amssymb}	
\usepackage{pdflscape}	
\usepackage{enumitem}   

\newcommand{\hii}{\mbox{H\,{\scshape ii}}}

\newcommand{\Ha}{\mbox{H$\alpha$}}
\newcommand{\Hb}{\mbox{H$\beta$}}
\newcommand{\Hg}{\mbox{H$\gamma$}}

\newcommand{\oiiiUVa}{\mbox{O\,{\scshape iii]}}\,$\lambda$1661}
\newcommand{\oiiiUVb}{\mbox{O\,{\scshape iii]}}\,$\lambda$1666}
\newcommand{\oiia}{\mbox{[O\,{\scshape ii]}}\,$\lambda$3727}
\newcommand{\oiib}{\mbox{[O\,{\scshape ii]}}\,$\lambda$3729}
\newcommand{\oii}{\mbox{[O\,{\scshape ii]}}\,$\lambda$3727,29}
\newcommand{\oiiia}{\mbox{[O\,{\scshape iii]}}\,$\lambda$4959}
\newcommand{\oiiib}{\mbox{[O\,{\scshape iii]}}\,$\lambda$5007}
\newcommand{\oiiiauroral}{\mbox{[O\,{\scshape iii]}}\,$\lambda$4364}

\newcommand{\nii}{\mbox{[N\,{\scshape ii]}}\,$\lambda$6585}
\newcommand{\niiauroral}{\mbox{[N\,{\scshape ii]}}\,$\lambda$5755}

\newcommand{\ciiia}{\mbox{[C\,{\scshape iii]}}\,$\lambda$1907}
\newcommand{\ciiib}{\mbox{C\,{\scshape iii]}}\,$\lambda$1909}

\newcommand{\neiii}{\mbox{Ne\,{\scshape iii}}\,$\lambda$3869}

\newcommand{\siia}{\mbox{[S\,{\scshape ii}]}\,$\lambda$6717}
\newcommand{\siib}{\mbox{[S\,{\scshape ii}]}\,$\lambda$6731}
\newcommand{\sii}{\mbox{[S\,{\scshape ii}]}\,$\lambda$6717,31}

\newcommand{\Te}{T$_\textrm{e}$\,}

\newcommand{\Ne}{n$_\textrm{e}$\,}
\newcommand{\ext}{E(B-V)\,}

\title[Metallicity Diagnostics at z$\sim$2]{Testing strong line metallicity diagnostics at z$\sim$2}

\author[Patr\'icio et al.]{
V. Patr\'icio,$^{1}$\thanks{E-mail: vera.patricio@dark-cosmology.dk}
L. Christensen,$^{1}$
H. Rhodin$^{1}$, R. Ca\~nameras$^{1}$ and M. A. Lara-L\'opez$^{1}$ \\
$^{1}$ Dark Cosmology Centre, Niels Bohr Institute, University of Copenhagen, Juliane Maries Vej 30, 2100 Copenhagen, Denmark\\
}

\date{Accepted XXX. Received YYY; in original form ZZZ}

\pubyear{2018}

\begin{document}
\label{firstpage}
\pagerange{\pageref{firstpage}--\pageref{lastpage}}
\maketitle

\begin{abstract}High-$z$ galaxy gas-phase metallicities are usually determined through observations of strong optical emission lines with calibrations tied to the local universe. Recent debate has questioned if these calibrations are valid in the high-$z$ universe. We investigate this by analysing a sample of 16  galaxies at $z\sim2$ available in the literature, and for which the metallicity can be robustly determined using oxygen auroral lines. The sample spans a redshift range of $1.4<z<3.6$, has metallicities of 7.4-8.4 in 12+$\log$(O/H) and stellar masses $\sim$10$^{7.5-11}$ M$_\odot$. We test commonly used strong line diagnostics (\emph{R23}, \emph{O3}, \emph{O2}, \emph{O32}, \emph{N2}, \emph{O3N2} and \emph{Ne3O2}) as prescribed by four different sets of empirical calibrations, as well as one fully theoretical calibration. We find that none of the strong line diagnostics (or calibration set) tested perform consistently better than the others. Amongst the line ratios tested, \emph{R23} and \emph{O3} deliver the best results, with accuracies as good as 0.01-0.04 dex and dispersions of $\sim$0.2 dex in two of the calibrations tested. Generally, line ratios involving nitrogen predict higher values of metallicity, while results with \emph{O32} and \emph{Ne3O2} show large dispersions. The theoretical calibration yields an accuracy of 0.06 dex, comparable to the best strong line methods. We conclude that, within the metallicity range tested in this work, the locally calibrated diagnostics can still be reliably applied at z$\sim2$. However, we caution that for $\sim60\%$ of our sample the observed \emph{R23} line ratios were out of the range of applicability of some of the calibrations tested. 
\end{abstract}




\section{Introduction}

One of the fundamental steps to understand the processes of galaxy growth and evolution is to robustly determine the chemical enrichment of galaxies over cosmic time. Gas-phase oxygen abundances (hereafter \textit{metallicity}) can be determined from a range of diagnostics tools applied to emission lines. Generally, these tools rely on strong optical emission lines and on empirical abundance diagnostics, such as the \emph{R23} \citep{Pagel1979}, \emph{O3N2} and \emph{N2} \citep{Pettini2004} ratios. It has long been known that these independently calibrated diagnostics are not compatible with each other, differing up to $\sim$0.15 dex \citep{Kewley2008}, which translates in a change of the mass-metallicity relation y-intercept of up to 0.7 dex in the local Universe \citep{Kewley2008} and about 0.3 dex at z$\sim0.6$ \citep{Maiolino2008}. 

Efforts have been made to derive consistent calibrations, but these studies rely mainly on local or low-redshift samples \citep[e.g.][]{Izotov2006,Pettini2004}. On the high-$z$ regime, there are concerns that strong-line diagnostics such as \emph{N2} do not deliver reliable metallicities, since \emph{N2} is sensitive to the variation of the N/O ratio and to the ionisation parameter, and there are strong indications that ionisation parameter change through cosmic time \citep{Steidel2014,Kewley2015}. The effort to validate the local metallicity diagnostics has been extended up to $z\sim1$ \citep{Ly2016,Jones2015}, but considering the number of surveys currently measuring galaxy metallicities at $z\sim2$ with KMOS \citep{Wisnioski2015} and MOSFIRE \citep{Steidel2014}, and the upcoming JWST spectra of high-redshift galaxies, it is timely to expand the metallicity diagnostics calibration to even higher redshifts.

The most direct method to determine metallicity relies only on atomic physics to calculate abundances. The intensity of an emission line depends on the abundance of the ion that it originates from, which we wish to determine, but also on the probability of that particular transition to occur, either through collisions or de-excitation, which is determined by atomic physics (see, for example, \citealt{Osterbrock1989}). This probability can be precisely calculated if the electron temperature and density are known. In this case, the several ion abundances of an element can be calculated and summed to obtain the total abundance. This method is called the \textit{direct method} ($T_e$) and it is the base of most empirical calibrations.

The electron temperature can be determined using line ratios involving auroral emission lines such as \oiiiauroral\, or \niiauroral\, (e.g. \citealt{Izotov2006}). These auroral lines become increasingly faint in lower-temperature \hii\, regions in high-mass (log $M_*>10.0$) and high-metallicity ($Z/Z_{\odot}>0.5$) galaxies, that are most frequently targeted in surveys. Conversely, low-mass galaxies have higher ionisation parameters, higher temperatures and UV lines become increasingly bright compared with more massive galaxies, but the intrinsic faintness of low-mass galaxies makes them challenging targets. This makes the detection of auroral lines very challenging, even in the local Universe. For example, of the original 1 200 000 galaxies from SDSS DR7, only 400 ($\sim$0.03\%) had detections \citep{Pilyugin2010}. At high-$z$, the detection of these lines becomes increasingly challenging and time consuming even on the largest aperture telescopes and the direct method is rarely used.\footnote{It is also possible to derive direct abundances from metal recombination lines, however, these are even fainter than auroral lines, and hardly used in extragalactic studies.} By targeting strongly lensed sources, which can have their flux magnified by a foreground massive galaxy cluster or massive galaxy up to a factor of $>$10, we can examine galaxies that are intrinsically fainter than those studied in flux-limited surveys. 

Strongly lensed galaxies at $z\sim2$ are still relatively rare, but observational efforts in the past decade have resulted in a sample of about two dozens of objects where auroral lines can be detected \citep[e.g.][]{Yuan2009,Dessauges-Zavadsky2010,Christensen2012a,Christensen2012b,James2014,Steidel2014}. Using these literature data, we assembled a sample of 16 galaxies at $z\sim2$, mostly lensed, where metallicities can both be derived through the direct and strong line methods, and use this sample to test the validity of commonly used strong line diagnostics. 

This paper is structured as follows. We describe our sample in Section~\ref{sec:data}. In Section~\ref{sec:electron_density}, we homogeneously re-derive the electron temperature for the entire sample. In Section~\ref{sec:direct_metalicity}, we derive the oxygen abundance based on the retrieved electron temperatures applying the direct method, and in Section~\ref{sec:empirical} and \ref{sec:theo} we compare these results with what is obtained using commonly used empirical and theoretical calibrations and finish with a summary and conclusions in Section~\ref{sec:conclusions}.

We adopt a solar metallicity of $12+\log$(O/H) = 8.69 \citep{AllendePrieto2001} throughout this work. All quantities mentioned in this paper have been corrected for the gravitational magnification factors available in the literature (see references in Tab~\ref{tab:sample}).

\section{Sample}
\label{sec:data}
 
We selected galaxies at $z>1$, available in the literature, for which the electron temperature can be determined either from the \oiiiUVb\, or the \oiiiauroral\, auroral lines. We also required that at least one Balmer ratio be available, in order to measure the intrinsic reddening and to de-redden emission line fluxes. The only exceptions to this are the Lynx arc \citep{Fosbury2003,Villar-Martin2004}, and SMACS2031 \citep{Christensen2012b,Patricio2016}, for which lower transition Balmer lines (\Ha, \Hb\, or \Hg) are not available to derive reddening. However there is evidence for very low dust content from SED fitting, so we also include these galaxies in this study. This resulted in a sample of 16 individual galaxies between $1.4<z<3.6$, all of them strongly lensed except three luminosity selected galaxies, plus a composite spectrum. We list all of these in Table~\ref{tab:sample}. 

\begin{table}
\caption{Sample used in this work. All objects are gravitationally lensed, with the exception of galaxies ID 9,10 and 11, which were broad band-selected. Redshift (z), stellar mass (M$_\star$, in M$_\odot$), star formation rates (SFR, in M$_\odot$/yr) as derived in each of the reference works. All masses and SFR were derived using the \citealt{Chabrier2003} IMF or converted from \citealt{Salpeter1955} IMF to \citealt{Chabrier2003} (see text).}
\label{tab:sample}
\tabcolsep=0.04cm
\begin{tabular}{|lllccc|}
\hline
ID & Object 	 & Reference 				& $z$ 	& log$_{10}$ M$_\star$ & SFR  \\
\hline\hline
1& CSWA20 			& \citealt{James2014}		& 1.433 & 10.3 	& 6   \\
2& Abell\_860\_359 & \citealt{Stark2014}		& 1.702 & 7.60 	& 30  \\
3& Abell\_22.3 	& \citealt{Yuan2009} 		& 1.703 & 8.5 	& 76  \\
4& RCSGA$^1$ & \citealt{Rigby2011} & 1.704	& 10.0 	& 106 \\
5& A1689\_31.1 	& \citealt{Christensen2012a,Christensen2012b} 	& 1.8 	& 7.7 	& 1\\
6& SMACS\_0304 	& \citealt{Christensen2012a,Christensen2012b}  	& 1.963 & 10.57 & 16  \\
7& MACS\_0451		& \citealt{Stark2014}			& 2.06 	& 7.49 	& 906 \\
8& COSMOS\_12805 	& \citealt{Kojima2017} 		& 2.159 & 9.24 	& 18 \\
9& BX660 			& \citealt{Erb2016}				& 2.174 & 9.73 	& 29  \\
10& BX74 			& \citealt{Erb2016} 				& 2.189 & 9.72 	& 58  \\
11& BX418 			& \citealt{Erb2016} 				& 2.305 & 9.45 	& 52  \\
12& S16-stack$^2$		&  \citealt{Steidel2016}			& 2.396 & 9.8 	& 29  \\
13& COSMOS-1908 	& \citealt{Sanders2016} 			& 3.077	& 9.3 	& 49  \\
14& Lynx arc$^3$ 		& \citealt{Villar-Martin2004} 	& 3.357 & 7.49 	& 54  \\
15& SMACS\_2031$^4$ 	& \citealt{Christensen2012a,Christensen2012b} 	& 3.5	& 9.16 	& 18 \\
16& SGAS\_1050 		& \citealt{Bayliss2014} 			& 3.625 & 9.5 	& 84  \\
\hline
\end{tabular}
\begin{flushleft}$^1$ RCSGA\_032727-132609; $^2$ Composite spectra; $^3$ See also \citealt{Fosbury2003}; $^4$ See also \citealt{Patricio2016}. \end{flushleft}
\end{table}

The sample spans a stellar mass range of 7.5-10.5 log$_{10}$ M$_\odot$ and a SFR range of a couple of solar masses per year up to almost 1000 M$_\odot$/yr. Most of these masses were calculated using the \citealt{Chabrier2003} initial mass function (IMF), with the exception of Abell\_860\_359 and the Lynx arc, where the \citealt{Salpeter1955} IMF was used. For these two objects, we correct the mass and SFR by dividing them by a factor of 1.8, to rescale these values to the \citealt{Chabrier2003} IMF. Several stellar population models were used, with the majority of works (13/16) using the \citealt{BC2003} models. The stellar mass of CSWA20 (ID 1) was not available in \citealt{James2014} and we calculate it from the 5-filter (ugriz) Sloan Digital Sky Survey photometry of image 1 in \citealt{Pettini2010}. We used the SED fitting code HyperZ \citep{Bolzonella2000} to determine the best fitting template spectra from a set of \citep{BC2003} models that use the \citealt{Chabrier2003} initial mass function. We fixed the redshift to that of the source, and computed the scaling factor for the best fit model, from which we can determine the stellar mass. The procedure was the same as outlined in \citealt{Christensen2012a}. We use the \citealt{James2014} magnification factor of $\mu$=11.5 to scale the observed magnitudes to their intrinsic (i.e. lensing corrected) values and compute the total stellar mass of the galaxy to be log M$_\star$ = 10.3$\pm$0.3 M$_\odot$. Overall, the sample spans a wide range of masses and SFRs, suitable to test metallicity calibrators at high-$z$ in different regimes.

\begin{figure}
\includegraphics[width=0.5\textwidth]{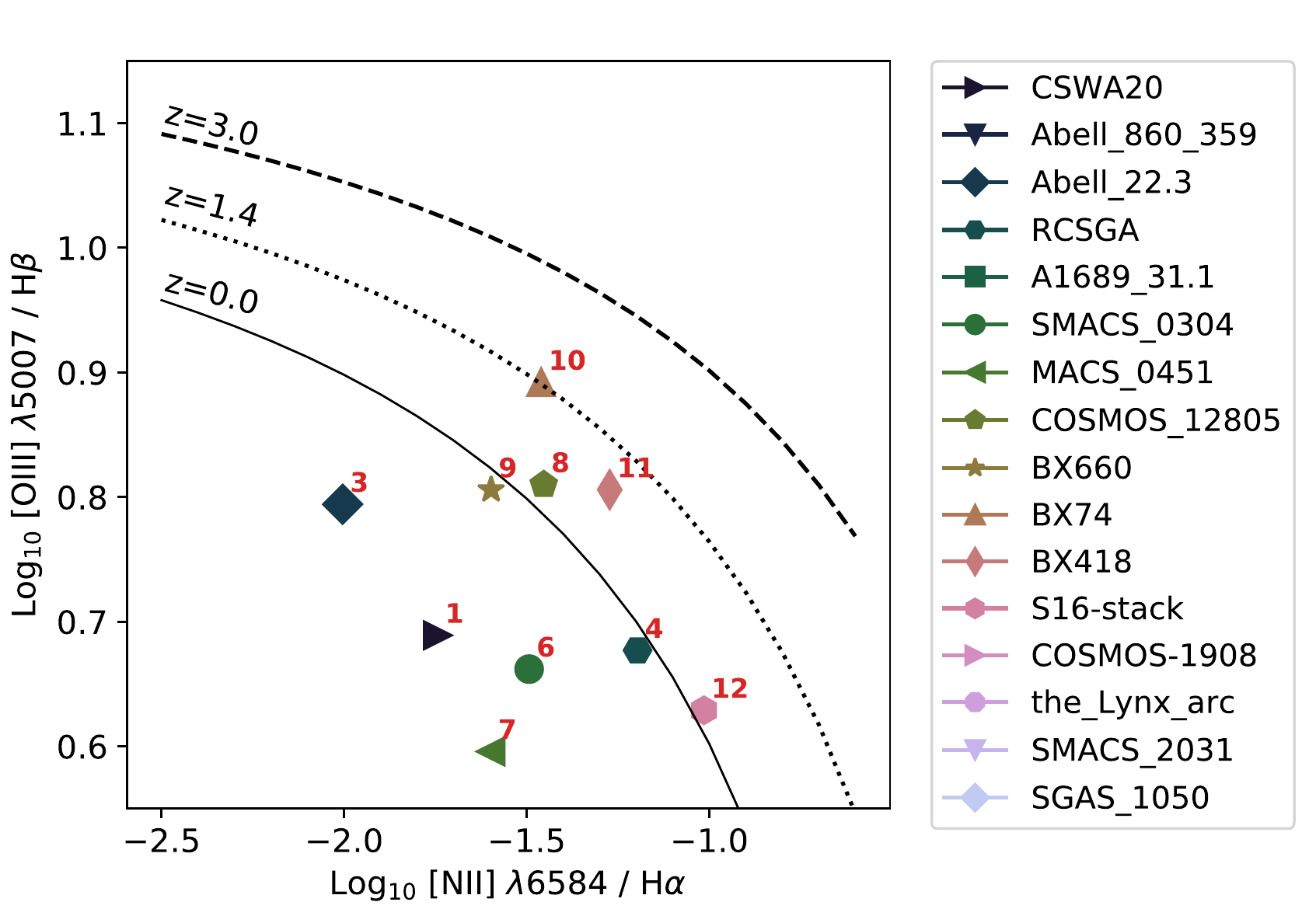}
\caption{BPT diagram with the classification line of \citet{Kewley2013} at z=1.4 and z=3.0 in dotted and full black lines, respectively. Line ratios were calculated with extinction corrected fluxes from Table~\ref{tab:fluxes}.}
\label{fig:main_sequence}
\end{figure}

None of the galaxies in the sample was identified as an AGN by previous works. We confirm this in Fig.~\ref{fig:main_sequence} (right panel) plotting our sample in the BPT diagram (\nii/\Ha\, vs \oiiib/\Hb, \citealt{Baldwin1981}) and comparing it with the classification criteria of \citealt{Kewley2013}, that separates AGN from star-forming dominated high-redshift galaxies. All the ten galaxies with the relevant lines available are consistent with being powered by star-formation and not AGN. Galaxies with ID 2, 5, 13, 14, 15 and 16 (see Table~\ref{tab:sample}) do not have the necessary lines to use this diagnostic plot, but for most galaxies other arguments can be used to rule AGN presence as unlikely. For Abell\_860\_359 (ID 2), Abel\_31.1 (ID 5) and SMACS\_2031 (ID 15), both the absence of \mbox{N\,{\scshape V}}\,$\lambda\lambda$1238,1242 and \mbox{He\,{\scshape II}}\,$\lambda$1640, and the low \mbox{C\,{\scshape IV}}\,$\lambda\lambda$1548,1550/\mbox{C\,{\scshape III}]}\,$\lambda\lambda$1907,1909 ratio when compared with typical narrow-line AGN values, make AGN presence not probable \citep{Stark2014,Christensen2012b}. \citealt{Sanders2016} argue that the low stellar mass and lack of strong peak emission in the central region of COSMOS-1908 (ID 13) make the presence of an AGN improbable. \citeauthor{Villar-Martin2004} claim that the Lynx arc (ID 14) emission lines ratios can be well described by models powered by star-formation and not an AGN. Finally, the SGAS\_1050 (ID 16) spectrum also does not contain emission lines usually associated with AGN emission, such as \mbox{N\,{\scshape V}}, and also no variation in the emission line widths is found in the data \citep{Bayliss2014}, removing AGN as the most likely dominant source of ionising photons. Overall, although it is not possible to absolutely exclude the presence of AGNs in our sample, it seems that star-formation is the main ionising source in these $z\sim2$ galaxies.

\section{Electron temperature and  Gas-phase redenning}
\label{sec:electron_density}

In this section, we re-derive the electron temperature and redenning corrections based on the line fluxes available in the literature. This ensures that our sample is treated homogeneously and that differences in \Te do not arise from the use of different atomic data. 

Before performing any analysis, we correct the spectra for Galactic extinction, using \citealt{Schlegel1998} dust map and the \citealt{Fitzpatrick1999} attenuation curve (assuming R$_V$ = 3.1). We used the {\sc dustmaps\footnote{http://dustmaps.readthedocs.io/en/latest/index.html} python} interface to recover the extinction value for each galaxy from the 2D dust map. This correction is nevertheless very small, in the order of 2\% for \Hb\ in most galaxies.

The first step to derive physical quantities from emission lines is to correct the intrinsic fluxes for dust extinction. The gas-phase redenning \ext can be derived comparing the theoretical values of Balmer lines ratios in the absence of dust with the observed ones. To compute the theoretical values, we make use of the \textit{getEmissivity} function in the {\sc PyNeb}\footnote{We use the "PyNeb\_17\_01" default atomic data set, only changing the \mbox{O\,{\scshape iii]}} data set to "o\_iii\_coll\_AK99.dat" from \citealt{Aggarwal}, in order to include level 6 transitions.} package \citep{Luridiana2015}. Balmer line ratios dependence on density is negligible for \Ne$<10^4$ cm$^{-3}$, as it is the case of our sample (see for example \citealt{Christensen2012a} where using the \oii\, doublet electron densities of \Ne=80-330 cm$^{-3}$ were derived). However, the ratios do depend on the electron temperature, although weakly. For this reason, we derive \ext\, and \Te\, iteratively. We use the reddened (i.e. observed, only corrected for magnification) fluxes to obtain a first guess on the electron temperature and density. Then, we compute the theoretical Balmer ratios in the absence of extinction for these conditions, and derive the extinction comparing these values with the observed ratios. The emission line fluxes are then dust corrected using the obtained \ext and the \citealt{Calzetti2000} extinction law, appropriate for star-forming galaxies. We then recompute the electron temperature from these dust corrected fluxes, and check if it is compatible with the initial value assumed to calculate the theoretical Balmer ratios. If not, the initial temperature is updated with the value obtained after the dust correction and we repeat the process until the difference between the electron temperature value used to calculate the Balmer ratios and the electron temperature obtained after the correction is less than 50 K.  

Electron temperature and density can be derived using adequate strong line ratios. These have to correspond to transitions from the same ion, so they do not depend on the total ion abundance. Moreover, to determine electron density, the two transitions also have to have their origin in atomic levels with very similar ionisation energies, and be mainly populated via collisional excitation \citep[e.g.][]{Osterbrock1989}. Collisional excitation strength depends both on the electronic density and temperature, but by selecting close energy levels the dependence on temperature is negligible. Conversely, by selecting levels with very different excitation energies, the dependence on density is minimized, and the electron temperature can be determined. The electron density and temperature can then be calculated using the observed ratios and the adequate atomic transition probabilities (see \citealt{Shaw1995} for a description of the basic equations used in the atomic transition models). We use the {\sc PyNeb} package \citep{Luridiana2015} to perform these calculations.

With the emission lines available in our sample, we can derive the electron temperature using two ratios: (\oiiiUVa+\oiiiUVb) / \oiiib\, and \oiiiauroral / \oiiib. If the two diagnostics are available, we make use of the \oiiiauroral\, ratio, since it is less sensitive to dust correction. 

The electron density can be derived using ratios of strong emission lines in the optical, such as \oiia/\oiib\, and \siia/\siib. However, \sii\, is not available for most of the sample and only in seven galaxies is the \oii\, doublet resolved. For this reason, we mostly derive densities using fainter UV lines, either \oiiiUVa\, and \oiiiUVb\, or \ciiia\, and \ciiib\, (see Table~\ref{tab:Te_and_ext}). The densities derived using these two ratios is known to differ up to an order of magnitude, most probably because these two ions have different ionisation energies and likely arise in environments of different densities, with \oiiiUVb\, tracing lower ionisation regions, that typically have lower electron densities (e.g. \citealt{James2014}). In order to have consistent results, and since \ciiia\, and \ciiib\, are only available for 4 galaxies which also have \oiiiUVa\, and \oiiiUVb\, measurements, we make exclusive use of the latter to derive densities. Whenever possible, we derive electron temperature and density at the same time, using the \textit{getCrossTemDen} {\sc PyNeb} function. If no density diagnostic is available, we assume a density of 100 cm$^3$, appropriate for \hii\, regions, and calculate the electron temperature\footnote{The electron temperature depends very weekly of the electron density, being practically constant for electron densities from 100 cm$^{-3}$ up to 1000 cm$^{-3}$.} using the \textit{getTemDen} {\sc PyNeb} function. 

We estimate the uncertainties of the derived quantities -- \Te, \ext, and \Ne -- from Monte Carlo simulations, varying the line fluxes assuming a Gaussian distribution with $\sigma$ equal to the associated uncertainties and centred in the measured fluxes. The values of electron temperature, density, redenning are shown in Table~\ref{tab:Te_and_ext} and correspond to the median value of the 500 Monte Carlo simulations of each galaxy. The lower and upper uncertainties correspond to the 16th and 84th percentiles of the same simulations. The dust corrected fluxes and their uncertainties are listed in Table~\ref{tab:fluxes}. Unphysical extinctions (i.e. \ext < 0) were assumed to be zero and no correction was applied to the fluxes.

Finally, we compare the values of electron temperature derived here with the ones available in the literature, finding that they are broadly compatible within uncertainties, with mean absolute difference of $\sim$1300 K and a minimum and maximum difference of 31 and 4460 K, respectively (see Fig.~\ref{fig:Te}).

\begin{table*}
\caption{Electron temperature, density, extinction and \Te based metallicity (see text for details). Diagnostics used for each galaxy are listed in the second column: a) \oiiiUVa+\oiiiUVb)/\oiiib; b) \oiiiauroral/\oiiib; c) \oiia/\oiib; d) \ciiia/\ciiib. Electron density for galaxies without available diagnostics was assumed to be 100 cm$^{-3}$. Ionic abundances and metallicity were obtained following \citet{Izotov2006} prescriptions.} 
\label{tab:Te_and_ext}
\renewcommand*{\arraystretch}{1.1}
\tabcolsep=0.25cm
\begin{tabular}{|lccccccc|}
\hline
Name & Diag. & \Te & \Ne & \ext 																& 12+$\log$(O$^+$/H$^+$) & 12+$\log$(O$^{2+}$/H$^+$) & 12+$\log$(O/H)$_{(Te)}$ \\
     &       & (K) & (cm$^{-3}$) & (mag.) 														& 					  &						   &  \\
\hline\hline
CSWA20	&b,c	&11895$^{+2688}_{-2459}$	&386$^{+141}_{-104}$	&0.13$^{+0.08}_{-0.08}$	 &7.04$^{+0.54}_{-0.35}$	 &7.99$^{+0.39}_{-0.23}$	 &8.04$^{+0.40}_{-0.24}$	 \\ 
Abell\_860\_359	&a,c	&13160$^{+1005}_{-1156}$	&100 	&0.04$^{+0.03}_{-0.03}$	 &7.28$^{+0.14}_{-0.09}$	 &8.01$^{+0.13}_{-0.09}$	 &8.09$^{+0.13}_{-0.09}$	 \\ 
Abell\_22.3	&a	&25640$^{+4280}_{-4730}$	&100 	&0.28$^{+0.26}_{-0.28}$	 &7.18$^{+0.38}_{-0.12}$	 &7.36$^{+0.17}_{-0.10}$	 &7.60$^{+0.09}_{-0.06}$	 \\ 
RCSGA	&b	&$\leq$11870	&100 	&0.24$^{+0.11}_{-0.07}$	 &$\geq$7.74	 &$\geq$7.98	 &$\geq$8.18	 \\ 
A1689\_31.1	&b,c	&20498$^{+4695}_{-3583}$	&234$^{+207}_{-139}$	&0.29$^{+0.30}_{-0.29}$	 &6.73$^{+0.13}_{-0.03}$	 &7.41$^{+0.18}_{-0.16}$	 &7.48$^{+0.16}_{-0.10}$	 \\ 
SMACS\_0304	&a,c	&12493$^{+592}_{-486}$	&1292$^{+3938}_{-1013}$	&0.21$^{+0.03}_{-0.01}$	 &7.86$^{+0.24}_{-0.12}$	 &8.02$^{+0.08}_{-0.07}$	 &8.26$^{+0.13}_{-0.09}$	 \\ 
MACS\_0451	&b	&21725$^{+2613}_{-2529}$	&100 	&0.00$^{+0.03}_{-0.00}$	 &6.70$^{+0.07}_{-0.05}$	 &7.34$^{+0.14}_{-0.11}$	 &7.43$^{+0.12}_{-0.09}$	 \\ 
COSMOS\_12805	&a	&12592$^{+3785}_{-751}$	&100 	&0.00$^{+0.27}_{-0.00}$	 &7.67$^{+0.18}_{-0.17}$	 &8.03$^{+0.17}_{-0.20}$	 &8.19$^{+0.18}_{-0.19}$	 \\ 
BX660	&a	&16244$^{+3446}_{-1076}$	&100 	&0.00$^{+0.17}_{-0.00}$	 &6.91$^{+0.10}_{-0.09}$	 &7.83$^{+0.11}_{-0.15}$	 &7.87$^{+0.11}_{-0.14}$	 \\ 
BX74	&a	&14231$^{+6678}_{-1490}$	&100 	&0.10$^{+0.26}_{-0.10}$	 &7.13$^{+0.18}_{-0.18}$	 &8.00$^{+0.17}_{-0.28}$	 &8.06$^{+0.17}_{-0.27}$	 \\ 
BX418	&a	&14667$^{+3427}_{-1212}$	&100 	&0.00$^{+0.19}_{-0.00}$	 &7.00$^{+0.14}_{-0.13}$	 &7.87$^{+0.17}_{-0.18}$	 &7.92$^{+0.16}_{-0.17}$	 \\ 
S16-stack	&a,c	&12861$^{+433}_{-453}$	&496$^{+154}_{-142}$	&0.21$^{+0.02}_{-0.02}$	 &7.74$^{+0.06}_{-0.07}$	 &7.94$^{+0.06}_{-0.07}$	 &8.15$^{+0.06}_{-0.07}$	 \\ 
COSMOS-1908	&b,c	&14633$^{+1757}_{-1655}$	&714$^{+764}_{-436}$	&0.09$^{+0.29}_{-0.09}$	 &6.82$^{+0.14}_{-0.12}$	 &7.92$^{+0.14}_{-0.12}$	 &7.96$^{+0.14}_{-0.12}$	 \\ 
Lynx arc	&a,d	&17369$^{+390}_{-345}$	&16014$^{+27546}_{-10973}$	&0 &6.82$^{+0.33}_{-0.23}$	 &7.79$^{+0.03}_{-0.03}$	 &7.83$^{+0.06}_{-0.04}$	 \\ 
SMACS\_2031	&a,c	&16684$^{+387}_{-346}$	&525$^{+308}_{-269}$	&0	 &6.83$^{+0.05}_{-0.05}$	 &7.62$^{+0.02}_{-0.02}$	 &7.68$^{+0.03}_{-0.02}$	 \\ 
SGAS\_1050	&a,c	&10735$^{+259}_{-295}$	&615$^{+597}_{-346}$	&0	 &7.42$^{+0.06}_{-0.06}$	 &8.37$^{+0.04}_{-0.03}$	 &8.41$^{+0.04}_{-0.03}$	 \\ 
\hline
\end{tabular}
\end{table*}

\section{Direct Metallicity}
\label{sec:direct_metalicity}

\begin{figure}
	\includegraphics[width=0.48\textwidth]{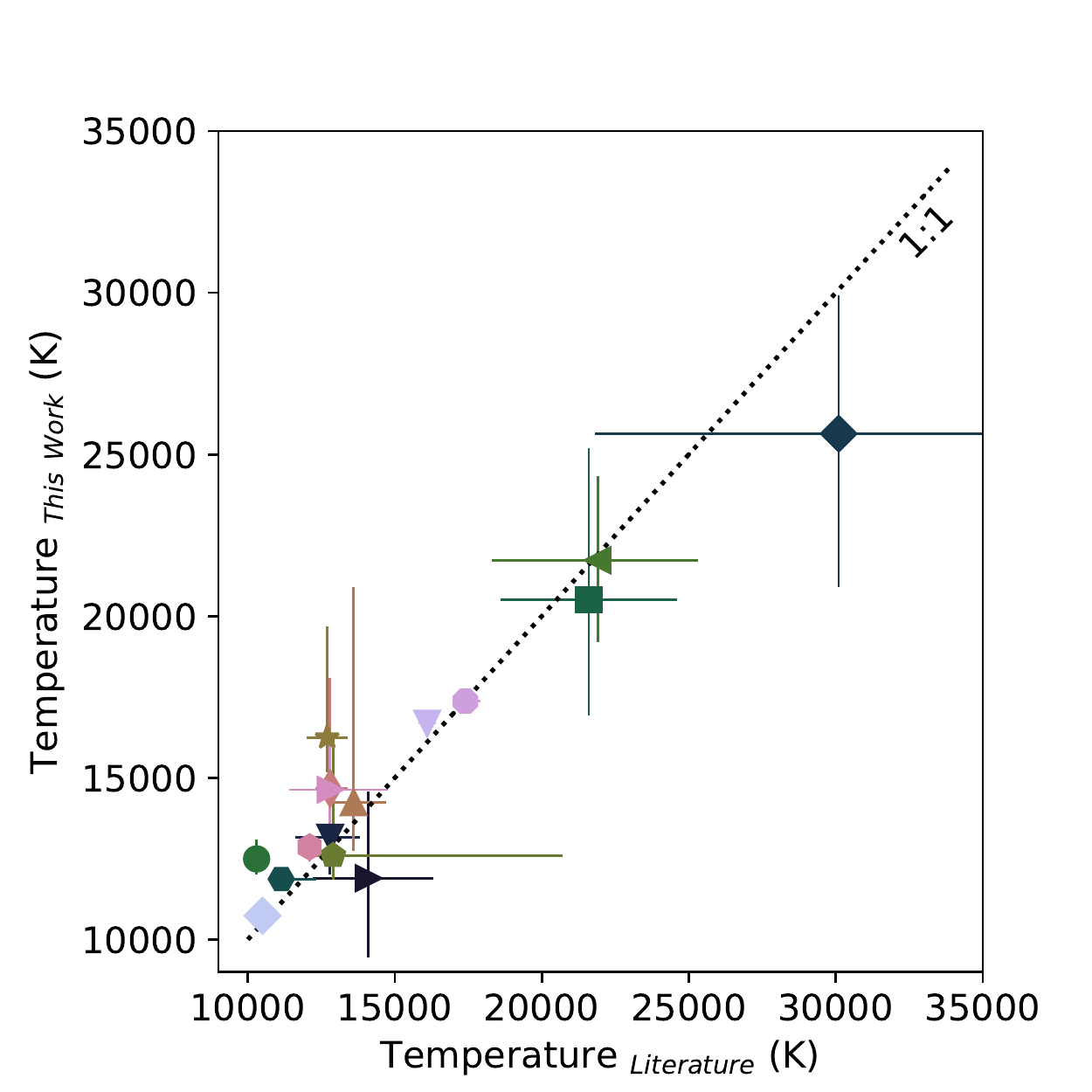}\\
    \includegraphics[width=0.48\textwidth]{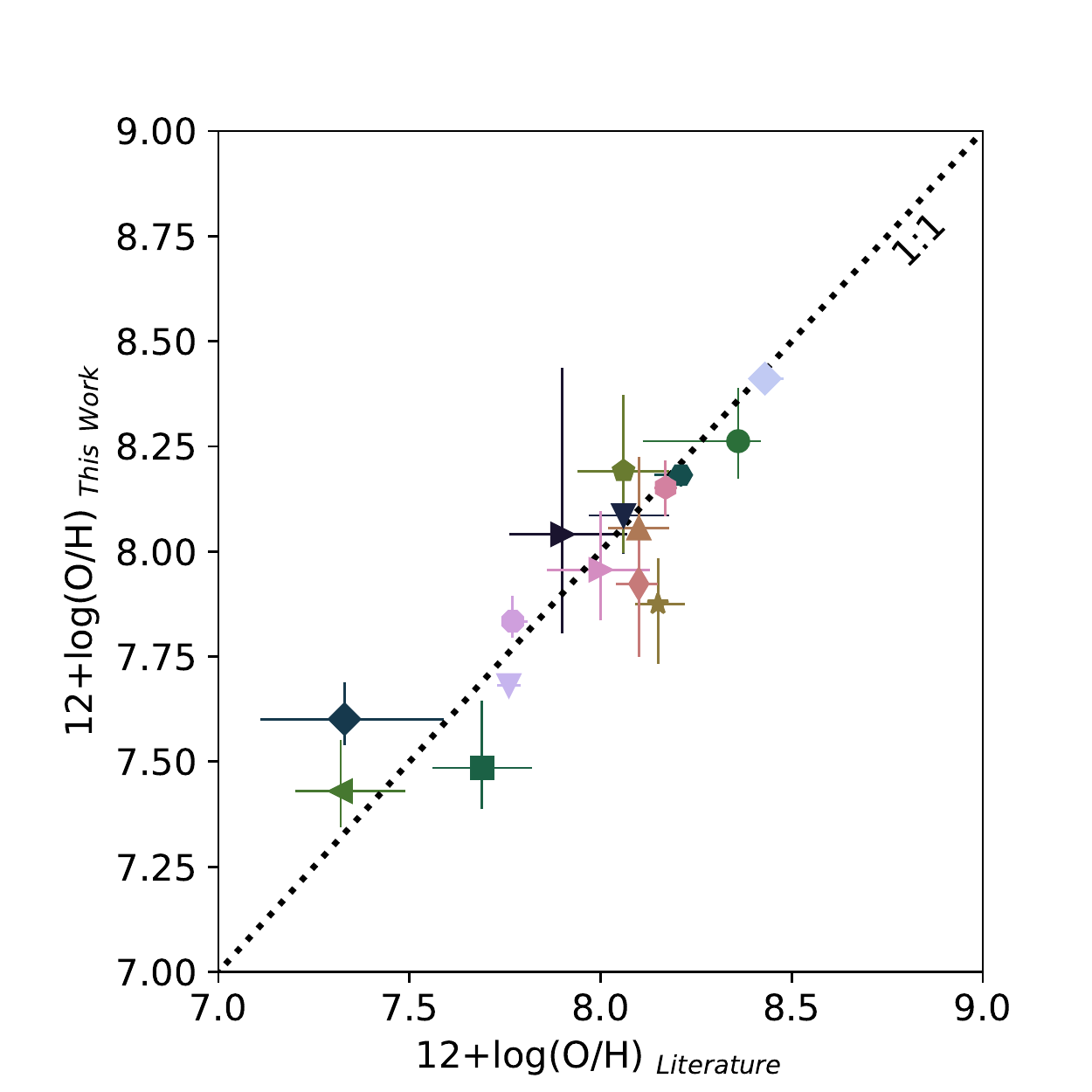}
    \caption{Comparison between electron temperature and direct metallicity from the literature and derived in this work. The 1:1 relation is plotted with a dotted line. We find a mean difference of $\sim$1300 K in electron temperature between the values derived in this work and the ones from literature, and a mean difference of $\sim$0.11 dex in metallicity. We use the same colours and markers as in Fig.~\ref{fig:main_sequence}.}
    \label{fig:Te}
\end{figure}

The total oxygen abundance can be calculated by adding the different ion abundances, that can in turn be derived if the electron temperature and density are known. Usually, only O$^+$/H$^+$ and O$^{2+}$/H$^+$ are accounted for, since higher ionisation states typically contribute to less than 1\% to the total abundances \citep{Izotov2006}. To calculate O$^+$/H$^+$ and O$^{2+}$/H$^+$ abundances we use the \textit{getIonAbundance} function of {\sc PyNeb}. This function assumes that the system is in equilibrium, i.e. the number of excitations and de-excitations is the same, and calculates the density of the ion relative to hydrogen based on the atomic transition probabilities, the atomic level populations (set by the electron density and temperature calculated in the previous section) and the observed line ratios (see \citealt{Luridiana2015} for details).

To determine O$^{2+}$/H$^+$, we use the derived electron temperatures and densities, and the extinction corrected fluxes of \oiiiauroral\, and \oiiib. The O$^+$ ion is formed in regions with lower electron temperature than O$^{+2}$, and we calculate the appropriate electron density for the O$^+$/H$^+$ ratio using \citealt{Izotov2006} equation 14 for low metallicities: 

\begin{equation}
t(\textrm{OII}) = -0.577 + t(\textrm{OIII}) \times (2.065 - 0.498\, t(\textrm{OIII}))
\end{equation}

\noindent with t(OII) = 10$^{-4}$T(OII) and t(OIII) = 10$^{-4}$T(OIII), and where T(OIII) is the temperature derived from \oiiiauroral\,, used to calculate the O$^{+2}$/H$^+$ ion abundance, and T(OII) the inferred temperature of O$^{+}$. The O$^+$/H$^+$ abundance is then estimated using T(OII) and again the \textit{getIonAbundance} function with extinction corrected fluxes of \oiia\, and \oiib. The total oxygen abundance is calculated by adding these two ion abundances. As before, errors are derived following a Monte Carlo technique, taking into account emission line fluxes, \Te and \Ne uncertainties. The results are listed in Table~\ref{tab:Te_and_ext} and compared with literature direct metallicities in Fig.~\ref{fig:Te}. We find a good agreement between the metallicity derived here and the literature values, with a mean difference of 0.11 dex (maximum of 0.27 dex and minimum of 0.01 dex), and without systematic differences with metallicity.

\section{Comparison with commonly used empirical calibrations}
\label{sec:empirical}

In this section, we compare the metallicities derived using the direct method with the results obtained using common strong line diagnostics, calculated using the extinction corrected fluxes listed in Table~\ref{tab:fluxes}. We examine strong line calibrations that make use of the following line ratios:

\begin{description}[align=right,labelwidth=1cm]
\item [\emph{R23}] (\oii+\oiiia+\oiiib)/\Hb 
\item [\emph{O3}] \oiiib/\Hb
\item [\emph{O2}] \oii/\Hb
\item [\emph{O32}] \oii/\oiiib 
\item [\emph{O3N2}] (\oiiib/\Hb) / (\nii/\Ha)
\item [\emph{N2}] \nii/\Ha
\item [\emph{Ne3O2}] \neiii/\oii
\end{description}

For each diagnostic, we calculate the metallicities applying the calibrations from \citealt{Maiolino2008}, \citealt{Curti2017}, \citealt{Jones2015} and \citealt{Bian2018} (hereafter M08, C17, J15 and BKD18), whenever available. Most calibrations are in the form of $\log_{10} R = \sum_{n} c_n x^n$, where $R$ is the line ratio, $c_n$ the calibration coefficients and $x^n$ the metallicity in 12+$\log$(O/H) - 8.69. The higher polynomial degree $n$ varies from 1 to 4, depending on the line ratio and calibration used. We compile all the calibrations and list them in four tables in Appendix~\ref{app:calib}. For each lines ratio value, we numerically find the metallicity that minimises the distance between the observed ratio and the one predicted by the appropriate strong line calibration ($\log_{10} R_{obs}-\log_{10} R$), using the {\sc python} package {\sc lmfit} \citep{lmfit}. The only exceptions are some calibrations of BKD18 (\emph{N2,O3N2,O32} and \emph{Ne3O2}), which are in the form $x = c_0 + c_1\log_{10} R$, and for which the metallicity can be derived directly without minimisation. We again use Monte Carlo simulations (varying line fluxes) to obtain the metallicity and upper and lower uncertainties. We do not include the intrinsic dispersion of the calibrations available both for \citealt{Curti2017} and \citealt{Jones2015} in our calculations of metallicity uncertainties, since we also wish to compare the dispersion obtained at $z\sim2$ with the one derived at lower redshifts.

To study the accuracy and precision of the strong line methods at $z\sim2$, we compute the difference between the metallicity derived using the strong line calibrations and the metallicity derived via the direct method (Sect.~\ref{sec:direct_metalicity}) and calculate both the mean ($\overline{x}$) and the standard deviation ($\sigma$) of the residuals. To test how sensitive these results are to our particular data set, since our sample is small, we utilise a bootstrapping technique, drawing 500 random samples with replacement of the available galaxies for each diagnostic. We describe the details of this process for each set of calibrations in the subsections below.

\subsection{\citealt{Maiolino2008} (M08)}

\begin{figure*}
	\includegraphics[width=\textwidth]{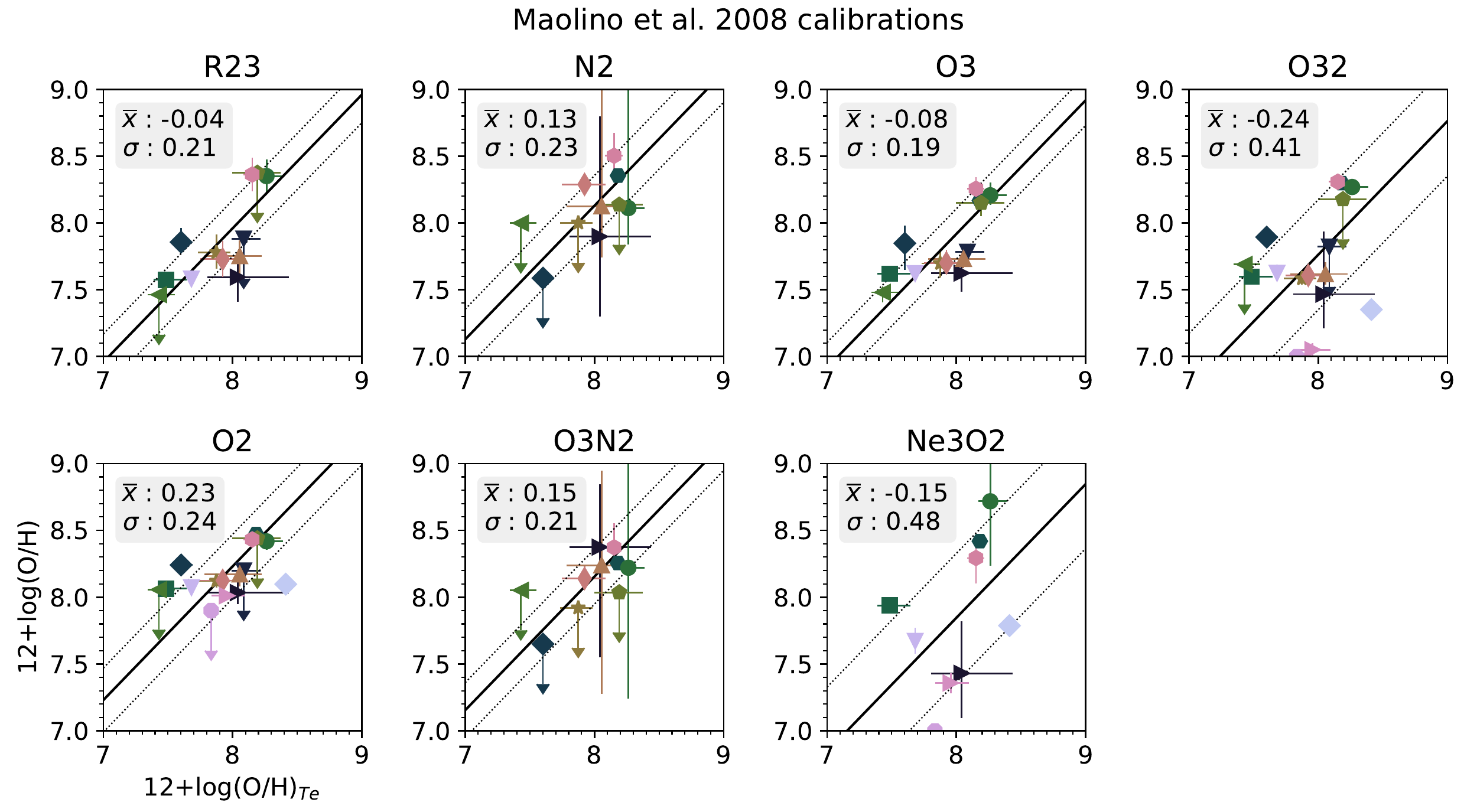}
    \caption{Comparison between metallicities derived using the \citet{Maiolino2008} calibrations (vertical axis) and the direct method (horizontal axis). Vertical uncertainties with arrows depict upper limits (where no uncertainties were available for one of the line fluxes). The mean and standard deviation of the residuals of the difference between these two values are shown in the upper left of each plot and are plotted in thick and dashed black lines, respectively. Objects have the same colours and markers as in Fig~\ref{fig:Te}.}
    \label{fig:maiolino}
\end{figure*}

M08 derive metallicity calibrations for the \emph{R23}, \emph{O3}, \emph{O2}, \emph{N2}, \emph{O32}, \emph{O3N2} and \emph{Ne3O2} ratios, using Sloan Digital Sky Survey (DR3 and DR4) galaxies. Note that in M08 \emph{O3N2} is the ratio between \oiiib/\nii, slightly different from the one used in other works and cited in the previous subsection. For the low metallicity range, the metallicities were derived using the direct method, while for the high metallicity range photoionisation models were utilised. These data extend from 7.0 up to 9.2 in 12+$\log$(O/H) and the calibrations were derived within this range (see M08 figure 5). Some of these ratios -- \emph{R23}, \emph{O3} and \emph{O2} -- are degenerate, with a high and low metallicity solution for each ratio value. If the direct method is not available, as it is almost always the case at high-$z$, the correct solution can be chosen, for example, based on the metallicity yield by non-degenerate calibrations: \emph{O2}3, \emph{O3\emph{N2}}, \emph{N2} or \emph{Ne3O2}. Alternatively, the mass-metallicity relation can also be used to choose the most likely metallicity solution.

When determining the metallicity of an object, all the available line ratios of the same set of calibrations should be used simultaneously, as done for example in M08, but since we aim at testing the accuracy of individual diagnostics applied to higher redshift galaxies we derive metallicities using each diagnostic independently. This means that we have to choose between one of the metallicity solutions when dealing with the three degenerate ratios. In our sample, the \emph{O32} diagnostic is available for all objects, so it can be used for this purpose. Although this is certainly not the case for all high-$z$ objects, we prefer to use this prior knowledge of the expected metallicity in order to reject the wrong metallicity branch, since our aim is to verify what is the accuracy of these diagnostics at $z\sim$2, and choosing the wrong branch would artificially increase the residuals.

For all diagnostics, we allow the solutions to vary between 12+$\log$(O/H) = 7 and 9, the range of validity of the M08 calibrations (based on their fig. 5). For each of the Monte Carlo simulations, we randomly pick the initial guess for the metallicity within this range. For the degenerate ratios however, the initial guess of the metallicity is based on the previously calculated \emph{O32} metallicity, which drives the solution towards the correct metallicity branch, although the allowed range of solutions still spans 7-9 in 12+$\log$(O/H). We obtained similar results when using direct method metallicity as the initial guess, instead of the \emph{O32} metallicity.

The \emph{R23}, \emph{O3} and \emph{O2} diagnostics present yet another issue: since they reach a local maximum in the middle of the considered metallicity range, there is a maximum value of the line ratio for which the calibrations can be used (the other, more linear, diagnostics also have maximum valid values of $\log$ R, but they are less problematic since they roughly correspond to the minimum or maximum of allowed metallicity). \emph{R23} reaches its maximum at 12+$\log$(O/H)=8.050 for log(\emph{R23})=0.938, \emph{O3} at 7.894 for log(\emph{O3})=0.746 and \emph{O2} at 8.702 for log(\emph{O2})= 0.5606. In our sample of 16 galaxies, 10 have \emph{R23} ratios higher to 0.938 (see Table~\ref{tab:ratios}), although, including uncertainties, this number reduces to 4 (ID 4,13,14,16). We notice that this perfectly natural, since the fit done to obtain the strong line calibrations minimises the distance to all data points, some galaxies will have ratios above the fit maximum (see M08 figure 5), and could possibly be solved by including the fit uncertainties. However, since these are not reported in M08, we avoid estimating metallicities outside of the validity range, we reject the 4 galaxies mentioned above and, for the remaining 6 problematic ones, we reject Monte Carlo simulations where the ratio is above the maximum of the calibration. For the \emph{O3} diagnostic, accounting for uncertainties, only 3 galaxies have ratios that are higher than 0.746, and all galaxies have \emph{O2} ratios lower than the maximum (see Table~\ref{tab:ratios}). 

Finally, we plot the results and compare them with the metallicities derived using the direct method in Fig.~\ref{fig:maiolino} and show the number of galaxies used in each comparison and mean and standard deviations (and respective uncertainties obtained using a bootstrapping technique) in Table~\ref{tab:stats}.

\subsection{\citealt{Jones2015} (J15)}

These calibrations were derived with a set of $z\sim0.8$ galaxies from the DEEP2 Galaxy Redshift Survey with available auroral lines. The authors derive calibrations for \emph{R23}, \emph{O2}, \emph{O3}, \emph{O32}, \emph{Ne3O2} and \emph{Ne3O3} with data that spans metallicities from 7.6 up to 9.0 in 12+$\log$(O/H). Comparing these calibrations with the ones obtained from local samples, J15 see no evolution with redshift, up to $z=0.8$. As in the M08 calibrations, \emph{R23}, \emph{O2} and \emph{O3} are degenerate and with maxima of 12+$\log$(O/H)=7.919 at log(\emph{R23})= 0.967, 12+$\log$(O/H)=7.845 at log(\emph{O3})=0.814 and 12+$\log$(O/H)=8.420 at log(\emph{O2})=0.561. We apply the same procedure here as for the M08 calibrations, searching for solutions within 7.6<12+$\log$(O/H)<9.0. The Ne3\emph{O3} diagnostic is extremely degenerate, with less than 0.2 dex variation in the line ratio for metallicities between 12+$\log$(O/H)=7.6 and 9.0, and we do not include it in the global comparison.

\begin{figure*}
	\includegraphics[width=0.8\textwidth]{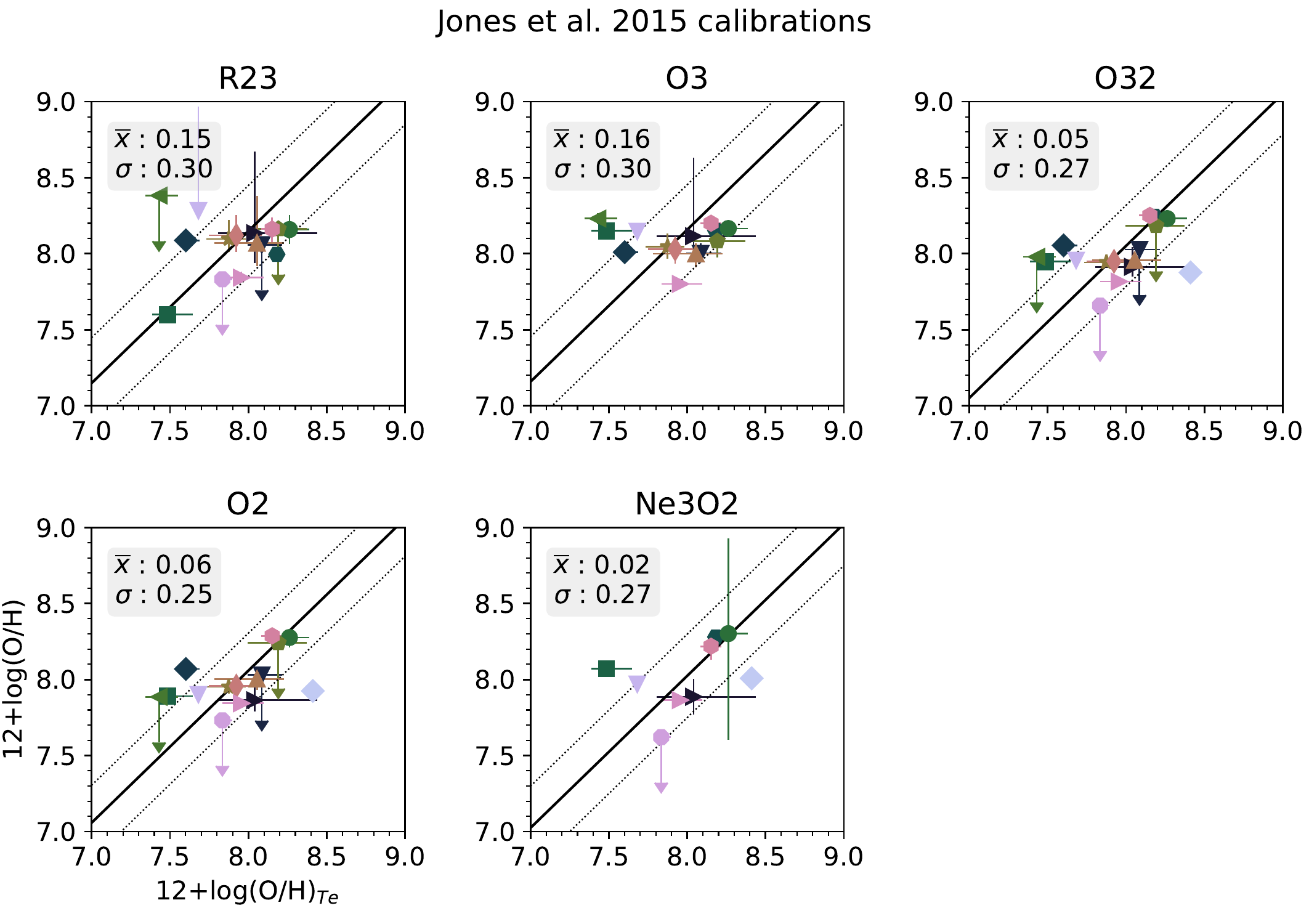}
    \caption{Comparison between metallicities derived using the \citet{Jones2015} calibrations and the direct method. See Fig.~\ref{fig:maiolino} for details.}
    \label{fig:jones}
\end{figure*}

We look for solutions between 7.6 and 8.6 in 12+$\log$(O/H) for all calibrations except for \emph{O3}, where the upper limit is set to 9.0, following J15 figure 9. The results are plotted in Fig.~\ref{fig:jones} and shown in Table~\ref{tab:stats}.

\begin{figure*}
	\includegraphics[width=0.8\textwidth]{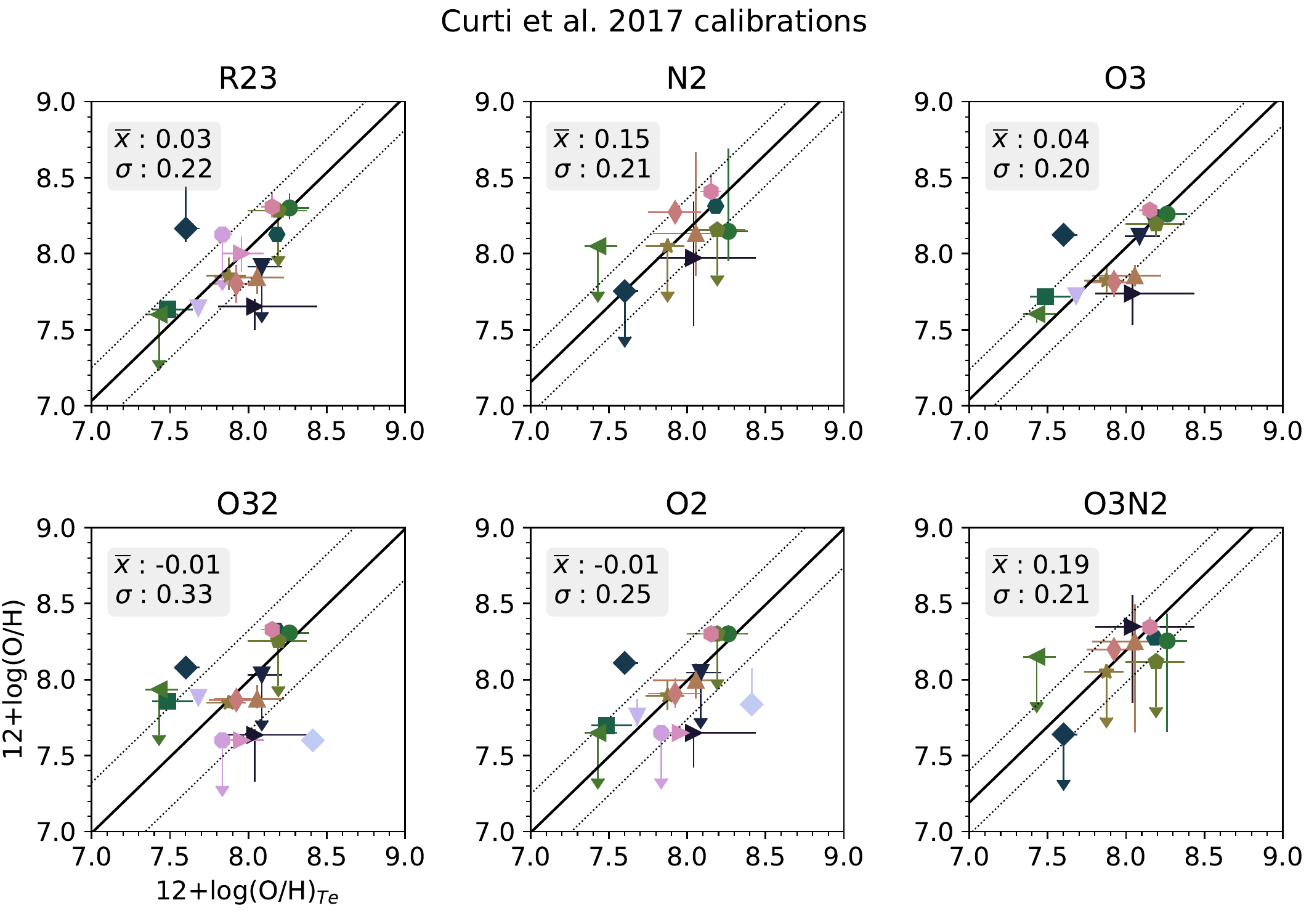}
    \caption{Comparison between metallicities derived using the \citet{Curti2017} calibrations and the direct method. See Fig.~\ref{fig:maiolino} for details.}
    \label{fig:curti}
\end{figure*}

\begin{figure*}
	\includegraphics[width=0.8\textwidth]{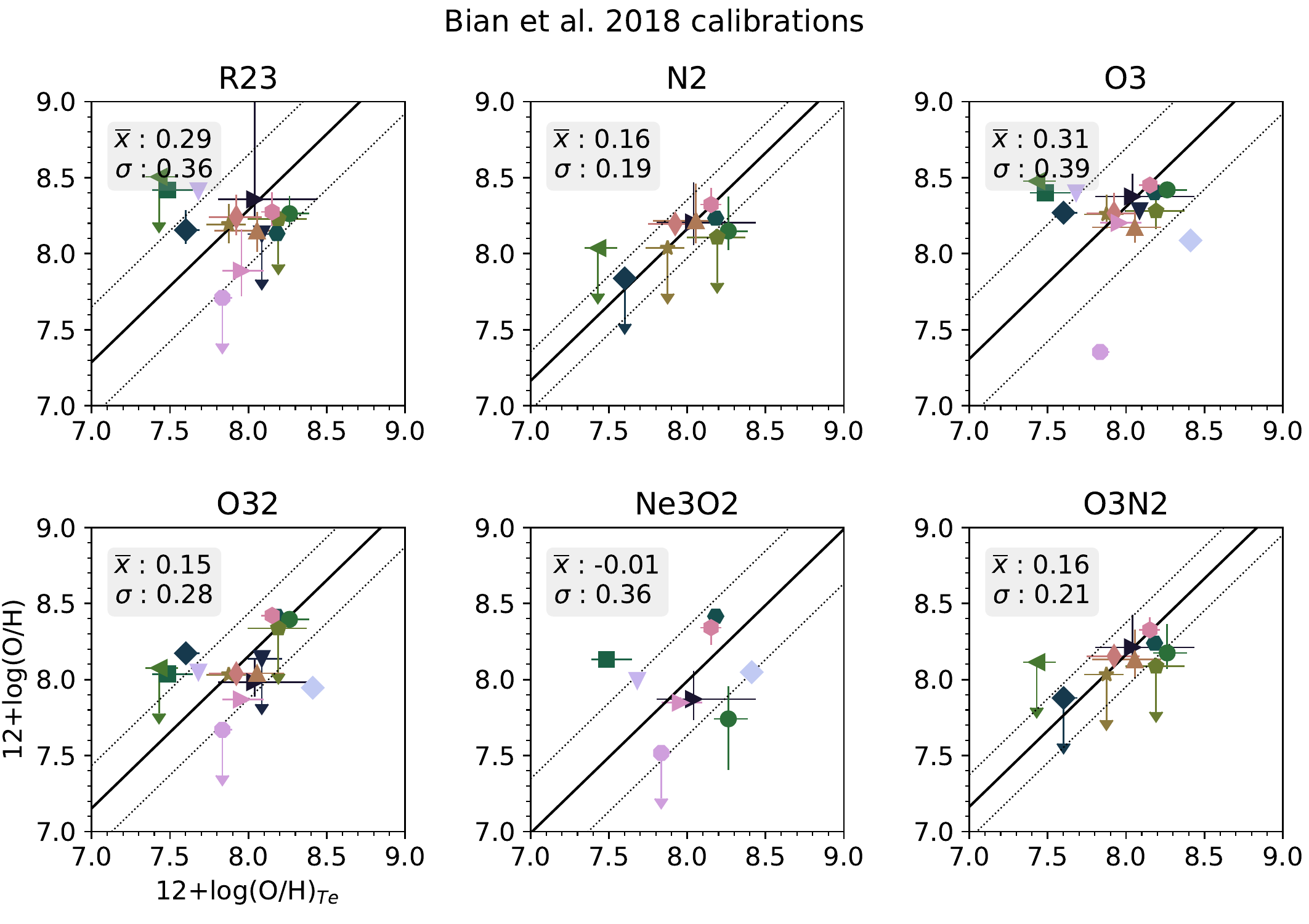}
    \caption{Comparison between metallicities derived using the \citet{Bian2018} calibrations and the direct method. See Fig.~\ref{fig:maiolino} for details.}
    \label{fig:bian}
\end{figure*}

\subsection{\citealt{Curti2017} (C17)}

These are fully empirical calibrations derived from a sample of 110000 Sloan Digital Sky Survey (DR7) galaxies. The authors stack galaxies accordingly to their strong line ratios in order to increase the signal to noise of the spectra, allowing robust detections of oxygen auroral lines, even in the highest metallicity ranges. C17 provide calibrations for the \emph{R23}, \emph{N2}, \emph{O3} (R3, in C17 nomenclature), \emph{O2} (R2), \emph{O32} and \emph{O3N2} diagnostics, and the metallicity ranges in which they can be applied, usually 7.6<12+$\log$(O/H)<8.85 (see C17 table 2). As in the previous calibrations, \emph{R23}, \emph{O2} and \emph{O3} are degenerate, although to a lesser extent, since the applicability range of these calibrations is more restricted. Indeed, \emph{R23} and \emph{O3} minimum valid metallicity is 8.4 and 8.3 in 12+$\log$(O/H), respectively. Within our sample, only one galaxy (SGAS\_1050, ID 16) has metallicity $\geq$8.3 12+$\log$(O/H), so in principle we should not be able to use these two diagnostics. Nevertheless, we still test \emph{R23} and \emph{O3} C17 calibrations, allowing the minimum metallicity to go down to 7.6 as the other diagnostics in C17, although we warn the reader that this is outside the applicability range given by C17. We calculate the metallicities following the same procedure detailed in the previous section and present the results in Fig.~\ref{fig:curti}.

\subsection{\citealt{Bian2018} (BKD18)}

Similarly to C17, this work also uses SDSS (DR7) stacked data to derive new metallicity calibrations from direct metallicity measurements. The main difference is the selection criteria of the stacked galaxies. The authors select 443 local galaxies that occupy the same BPT region as the z$\sim2.3$ star-forming galaxies analysed in \citealt{Steidel2014}. This  sample of local z$\sim2$ analogs is then stacked based on their \nii/\Ha\, ratios, producing 7 stacked spectra in which the \oiiiauroral\, line can be detected and used to calculate the direct metallicity.

BKD18 derive calibrations for \emph{R23}, \emph{N2}, \emph{O3}, \emph{O32} and \emph{O3N2} diagnostics. Only \emph{R23} and \emph{O3} are degenerate with local maxima at 12+$\log$(O/H) = 7.87 and $\log_{10}$(\emph{R23}) = 0.99 and 12+$\log$(O/H) = 7.77 and $\log_{10}$(\emph{O3}) = 0.98, respectively. Only SGAS\_1050 has an \emph{R23} ratio higher than this local maximum and all galaxies have \emph{O3} ratios lower than 0.98.

The stacked spectra have metallicities between 7.8 < 12 + $\log$(O/H) < 8.4, and does not extend in metallicity range to include the 3 lowest metallicity galaxies in our sample (see Table~\ref{tab:Te_and_ext}). Nevertheless, as for the previous calibrations, we also include these galaxies in the sample, allowing the metallicity solutions to be as low as 7 in 12+$\log$(O/H). The comparison between the metallicities obtained with the BKD18 calibrations and the direct metallicities is shown in Fig~\ref{fig:bian}.

\subsection{Strong line methods at $z\sim2$}

\begin{figure}
	\includegraphics[width=0.5\textwidth]{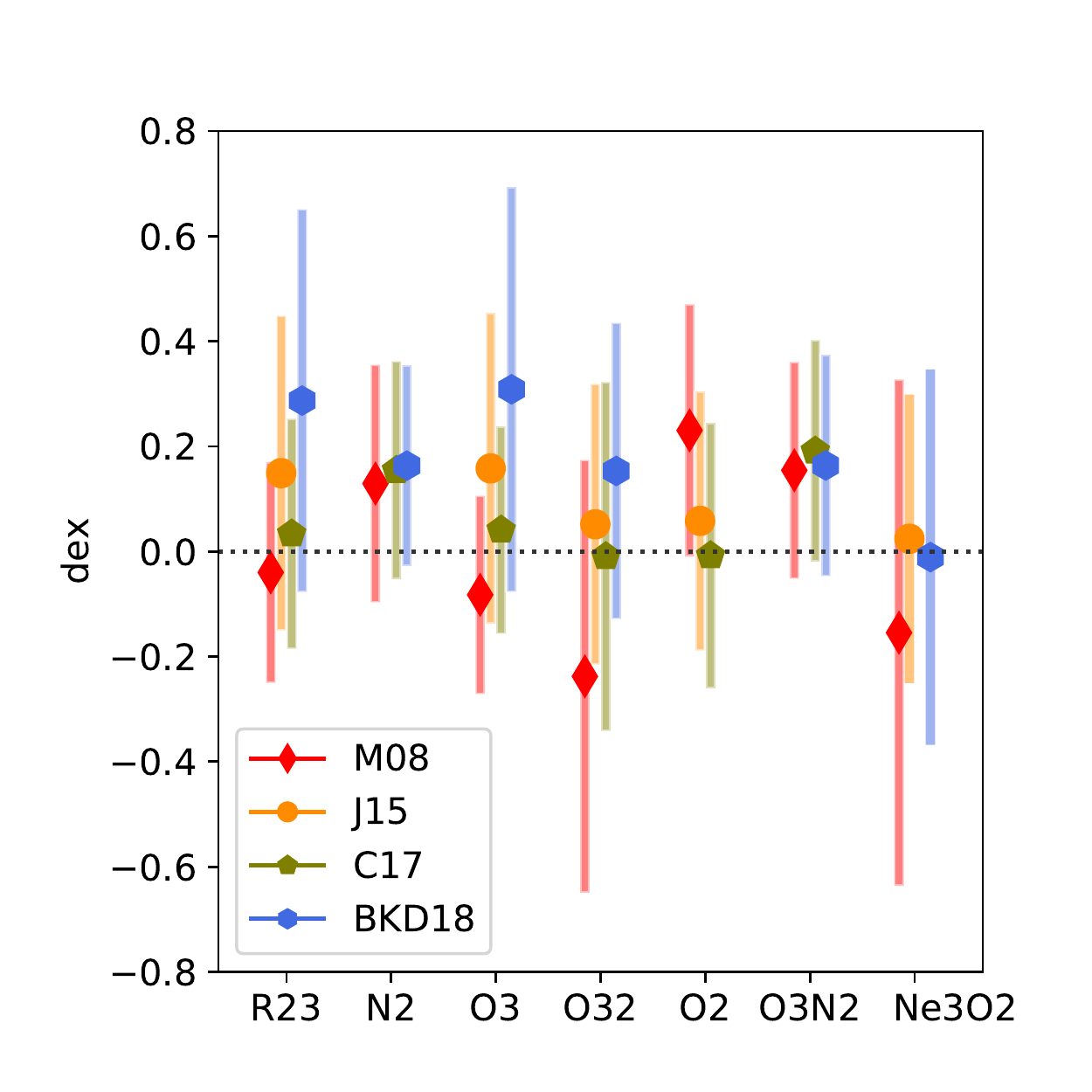}
    \caption{Summary of the accuracy and precision of the 7 strong line ratios tested here (in the horizontal axis) for the different empirical calibrations tested here (M08, J15, C17 and BKD18, in red, orange, green and blue, respectively). The mean difference between each ratio and the direct metallicity ($\overline{x}$) is given by the markers, and the standard deviation of the residuals ($\sigma_{z~2}$) is given by the vertical bars. Error bars on both quantities can be found in Table~\ref{tab:stats}.}
    \label{fig:all_calibs}
\end{figure}

We summarise our results in Figure~\ref{fig:all_calibs}, presenting the mean and standard deviation of the residuals between the metallicity derived using each diagnostic and calibration and the direct method.

We first analyse the accuracy and precision at $z\sim2$ of the different diagnostics within the same calibration set. For each diagnostic, we use the mean and the standard deviation of the residuals between the metallicity obtained using strong lines methods and the one derived using the \Te method (that we take to be the true metallicity). All values can be found in Table~\ref{tab:stats}.

For M08, the lowest residuals are found for \emph{R23}, with mean residuals of -0.04$\pm$0.06 dex. The second most accurate diagnostic is \emph{O3}, under-predicting the metallicity only by -0.08$\pm$0.06 dex, when compared with the direct method. The worst accuracy is obtained with \emph{O32} and \emph{O2}, with mean residuals of -0.24$\pm$0.11 and 0.23$\pm$0.06 dex, respectively. Most diagnostics have a precision (dispersion in the residuals) of about 0.2 dex, with the exceptions of \emph{O32} and \emph{Ne3O2}, that have residual $\sigma_{z\sim2}$ around 0.4 dex. J15 calibrations, the only ones tested here that were calibrated also using non-local galaxies (up to $z\sim0.8$), show an overall good agreement with the direct metallicities, with a maximum mean residual of $\sim$0.15 dex for \emph{R23} and \emph{O3}, and the remaining diagnostics with $\overline{x}\le$ 0.06 dex. The dispersion in the residuals of J15 are slightly higher than for M08 and C17, with $\sigma_{z\sim2}$ around 0.3 dex. Using C17 calibrations, the lowest mean residuals are found using the \emph{R23}, \emph{O2} and \emph{O32} diagnostics, even when utilised outside their applicability range. The dispersions are comparable to M08, varying between 0.21 and 0.33 dex. Finally, the \emph{Ne3O2} diagnostic has a low mean residual in the BKD18 diagnostics ($\overline{x}$=-0.01$\pm$0.11), although this value is higher for other diagnostics, typically at about 0.16 dex, but as high as 0.3 dex for \emph{R23} and \emph{O3}. The precision of these calibrations is also usually lower, with $\sigma_{z\sim2}$ typically around 0.3 dex and never lower than 0.19 dex, but comparable to the other calibrations tested.

Perhaps unsurprisingly, the dispersion at $z\sim2$ is higher than the dispersion found both by J15 and C17 (calculated from data that is mostly comprised by low redshift objects). For J15, the typical dispersion is $\sim0.28$ dex for high-$z$, while it varies between 0.06 and 0.23 dex for their studied sample. For C17 calibrations, we obtained dispersions between 0.2 and 0.3 dex, where in the local sample used to derive these calibrations this value varies between 0.07 and 0.26 dex. 

We find that no single line ratio systematically outperforms the others, neither does one set of calibrations. The \emph{R23} and \emph{O3} diagnostics in the M08 and C17 calibrations have both a good accuracy, with mean residuals very close or compatible with zero, and a good precision, with dispersions of about 0.2 dex. The \emph{O2} ratio also yields low mean residuals with the J15 and C17 calibrations, but has a higher dispersion (0.25 dex) than  \emph{R23} and \emph{O3}. \emph{N2} and \emph{O3N2} are systematically offset in all calibrations, generally predicting higher metallicities than the ones measured using the direct method. Since both diagnostics include nitrogen, this might possibly indicate an evolution of the nitrogen abundance relative to oxygen with cosmic time, although our analysis and sample size is not suited to study this evolution. Finally, both \emph{O32} and \emph{Ne3O2} have large dispersions with all calibrations, varying between 0.27 to 0.48 dex, which might result in metallicities off by a factor of $\sim$2-4 when applying these diagnostics at $z\sim$ 2. These calibrations primarily depend on the ionisation parameter,
their dependence on metallicity is mostly an indirect consequence
of a correlation between metallicity and ionisation parameter in most galaxies \citep{Kewley2002}. It is therefore not surprising that the dispersion obtained with these diagnostics is higher than with those directly depending on metallicity.

We remind nevertheless that these results were obtained with a small sample, from 16 objects \emph{O32} and \emph{O2} but as low as 9 for \emph{Ne3O2}, and a higher number of objects would be desirable to establish these results more robustly. Furthermore, we also probe a restricted metallicity range $\sim$7.4-8.4 12+$\log$(O/H) (see Fig.~\ref{fig:Te}), that does not fully cover the high metallicity part of M08 and J15 calibrations, that used data up to $\sim$9 12+log(O/H), where, for example the \emph{N2} diagnostic is likely to become problematic. In the case of C17 calibration, the metallicity range probed here is close to the applicability range given by the authors (7.6 < 12+$\log$(O/H) < 8.85, see their table 2), so this is less of an issue. However, most objects of our sample have metallicities between 8 and 8.5 in 12+$\log$(O/H), so it would be important to increase the number of low and high metallicity galaxies in this study in order to increase the robustness of these results, and to possibly look for trends in the residuals with metallicity.

\section{Comparison with P\'erez-Montero et al. 2014 calibrations}
\label{sec:theo}

\begin{figure}
	\includegraphics[width=0.5\textwidth]{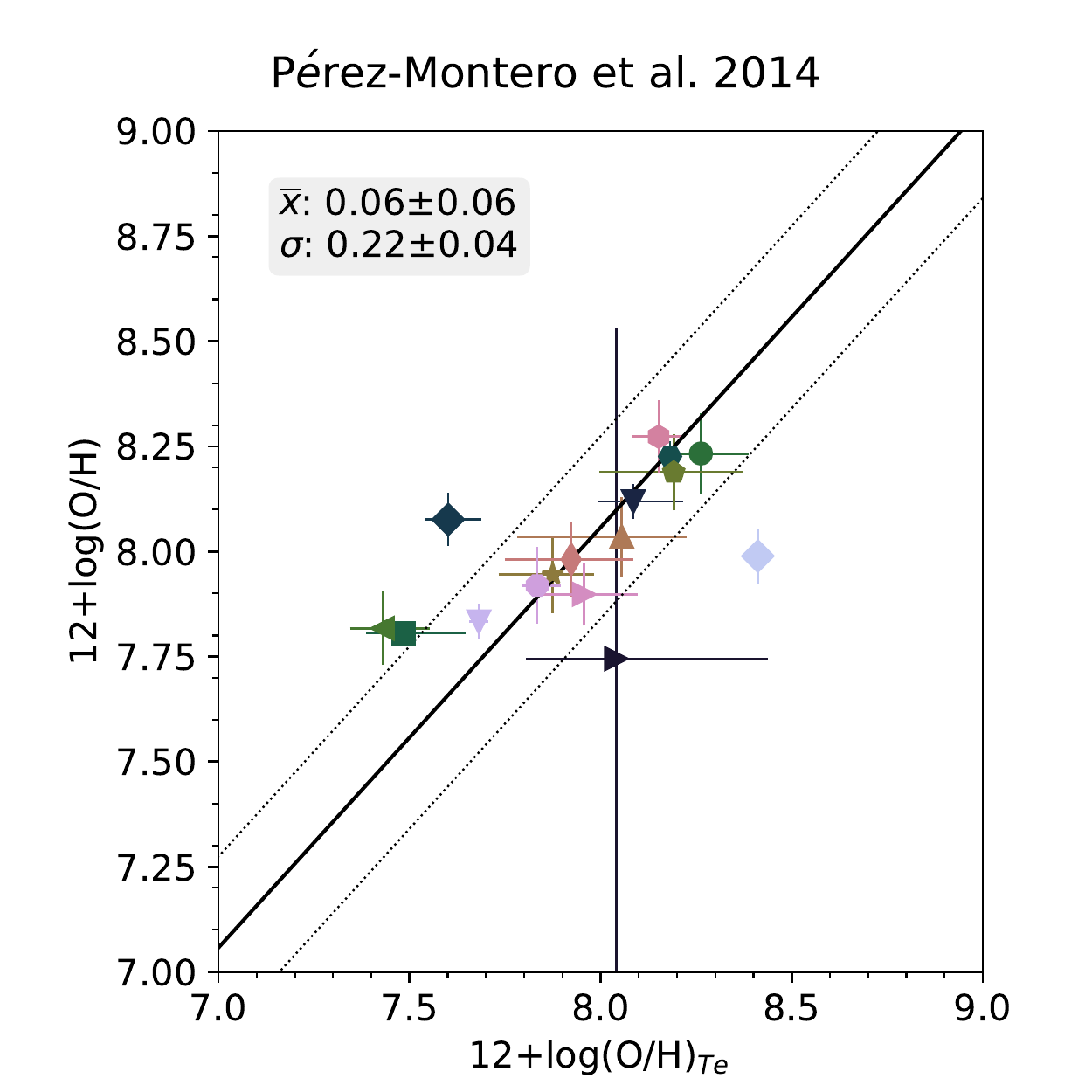}
    \caption{Comparison between metallicity derived using the \citealt{Perez-Montero2014} calibrations (including all strong emission lines available, but not \oiiiauroral) and the direct method. The same markers and colours as in Fig.~\ref{fig:maiolino} are used here.}
    \label{fig:perez}
\end{figure}

Contrary to the previous calibrations, mostly empirical, \citealt{Perez-Montero2014} (hereafter PM14) provide a set of fully theoretical calibrations. The authors provide a grid of photoionisation models, obtained using {\sc cloudy} \citep{cloudy} to calculate emergent spectra. The grid spans a large range of O/H ( 7.1 < 12+$\log$(O/H) < 9.1) and N/O abundances (0 < log(N/O) < -2 ) as well as ionisation parameters  (-4 <$\log(U)$ < -1). PM14 also provides a {\sc python} implementation of their procedure, {\sc hii-chi-mistry}\footnote{http://www.iaa.es/\~epm/HII-CHI-mistry.html}, that determines the best grid model for a given set of emission lines. Unlike the previously tested calibrations, this method makes use of all the emission ratios at the same time, finding the best model in the grid taking into account all line fluxes. 

PM14 test their method using a large sample of local objects where auroral lines are detected. When all the emission lines are utilised -- \oii, \oiiiauroral, \oiiib, \nii, \mbox{[S\,{\scshape ii]}}\,$\lambda$6717,6731 and \Hb\, -- they find a dispersion of 0.07 dex for 12+$\log$(O/H)<8.0 and 0.14 dex for 12+$\log$(O/H)>8.0 of their solutions when compared with the direct method. PM14 also test the scenario where the \oiiiauroral\,line is not detected, usually the case at high metallicity and at high-$z$ due to the faintness of the auroral line. They find that this yields poor results when all models in the grid are included, but limiting the search to a sub-sample of models in the grid determined in P14 (shown their figure 3), a much better agreement is obtained. When the \oiiiauroral\, line is not available, and using the set of empirically limited grids in $\log$ U (ionisation parameter), a dispersion of 0.16 dex for the low metallicity objects (12+$\log$(O/H)<8.0) and 0.05 dex at higher metallicities is found. If besides \oiiiauroral, the \nii\, line is not available, a set of log(N/O) empirically limited grids can be used, and a global dispersion of 0.19 dex is reached in the local sample (although with an overestimation of 0.26 dex of the oxygen abundances for 12+$\log$(O/H)<8.0 objects).

We investigate the results obtained with the PM14 method using all the strong emission lines available for each galaxy, but not the \oiiiauroral\ emission line.
In this case, the procedure automatically selects the empirically limited grids to be used. We use version 3 of {\sc hii-chi-mistry} with interpolated grids, with a resolution of 0.02 dex for 12+$\log$(O/H) and 0.025 dex for log(N/O), for all tests. The results and comparison with the $z\sim2$ sample are shown in Fig.~\ref{fig:perez}.

The results obtained have low mean residuals ($\overline{x}$ = 0.06$\pm$0.05 dex)  and also a reasonable low dispersion ($\sigma=0.22\pm0.04$), which is comparable to the best empirical calibrated diagnostics (e.g. \emph{R23} with the M08 and J15 calibrations or \emph{O32} with the C17 results).

\section{Summary and Conclusions}
\label{sec:conclusions}

We gathered a sample of 16 high-$z$ galaxies from $z\sim1.4$ up to $z\sim3.6$ where auroral lines are detected, with the aim of testing the accuracy and precision of the metallicity derived using strong line methods at $z\sim2$. Using line fluxes from literature, we homogeneously re-derive the oxygen abundances through the direct method, which relies on the determination of the electron temperature from auroral line ratios, and is considered the most accurate method available. 

We compare the direct metallicities to the metallicities derived using the strong line ratios \emph{R23}, \emph{O32}, \emph{O3}, \emph{O2}, \emph{N2},  and \emph{Ne3O2}. To do so, we test four different calibration sets from \citealt{Maiolino2008}, \citealt{Jones2015}, \citealt{Curti2017} and \citealt{Bian2018}. Since some of these commonly used diagnostics (\emph{R23}, \emph{O3} and \emph{O2}) are degenerate, this is, for the same ratio value there are two possible metallicity solutions, some a priori knowledge about the metallicity is needed in order to choose the correct branch. Here, we base our choice in the results obtained with \emph{O32}. Moreover, these degenerate diagnostics have a local maximum in the metallicity range considered, so there is also a maximum ratio for which the calibrations can be formally applied. In our sample of 16 galaxies, 10 presented \emph{R23} ratios higher than this allowed maximum. This may bring some extra uncertainty when using the \emph{R23} ratio to derive metallicities close to 12+log(O/H)=8.0.

We did not find a line ratio that delivered better results irrespectively of the calibration used, neither a calibration set that outperformed the others in all diagnostics. However, there are some combinations that deliver better results, which have both low mean residuals and dispersion. We summarise these results below:

\begin{itemize}
\item \emph{R23} and \emph{O3} with the M08 and C17 calibration have both mean residuals very close or compatible with zero, and small dispersions ($\sim$ 0.2 dex);
\item \emph{O2} with the J15 and C17 calibrations has comparable low mean residuals (0.06 and -0.01 dex, respectively), although a slightly higher dispersion ($\sim$ 0.25 dex);
\item \emph{N2} and \emph{O3N2} have high and positive mean residuals with all the calibrations tested. This might be due to an evolution with redshift of the abundance of nitrogen relative to oxygen;
\item \emph{O32} and \emph{Ne3O2} have large dispersions with all calibrations, varying between 0.27 to 0.48 dex. Since these diagnostics are primarily dependent on the ionisation parameter, this variability is somehow expected;
\item The theoretical calibration of \citealt{Perez-Montero2014} achieve comparable results to the best line ratios diagnostics using all available bright emission lines.
\end{itemize}

There are some caveats in this study. First, the number of galaxies at high-$z$ for which the auroral line is detected is still small, and a bigger sample should be analysed in order to robustly confirm our results. Moreover, although we cover metallicities as low as 12+log(O/H) = 7.4, the higher metallicity regime, 12+log(O/H)>8.4, was not tested in this study, which might be an issue when considering the most luminous, massive, and hence more metal rich, galaxies. However, the metallicity range examined in this work mostly covers the regime of $z\sim2$ main sequence galaxies that currently have metallicity estimates. As an example, the highest metallicity of the MOSDEF Survey sample of 87 star-forming galaxies is about 8.7 in 12 +log(O/H) \citep{Sanders2015}, approximately the same as the smaller sample of $z\sim2.2$ galaxies in \citealt{Maiolino2008}. Therefore, we find that the mass and metallicity range covered in this work is adequate to confirm that most strong line methods can still be applied with reasonable confidence at $z\sim2$. We plan to increase the sample size and expand the mass regime of galaxies at $z\sim2$ with auroral line detections in order to confirm these results. This will also allow to explore for the first time the mass-metallicity at $z\sim2$ employing direct metallicity estimations only, that do not depend on the local calibrations.

\section*{Acknowledgements}

We thank Michael Maseda and Charles Steinhardt for useful comments about high-$z$ flux measurements and statistics. We also thank the anonymous referee for providing useful comments that improved this work.

VP, LC, RC and HR are supported by the grant DFF - 4090-00079. This research made use of {\sc astropy}, a community-developed core Python package for Astronomy \citealt{astropy}. This research has made use of NASA's Astrophysics Data System. 

The full analysis presented in this paper is available in \url{https://github.com/VeraPatricio/MetCalibration.git}.



\bibliographystyle{mnras}
\bibliography{bibliography} 



\appendix

\section{Metallicity calibrations}
\label{app:calib}

Here we list the metallicity calibrations used in this work. All calibrations from \citealt{Maiolino2008}, \citealt{Jones2015} and \citealt{Curti2017} are given as $\log_{10} R = \sum_{n} c_n x^n$, where $R$ is the line ratio, $c_n$ the calibration coefficients and $x^n$ the metallicity in 12+$\log$(O/H) - 8.69. For the \citealt{Bian2018} calibrations, only \emph{R23}, \emph{O3} and \emph{Ne2O2} are given in function of the metallicity, while \emph{O32}, \emph{N2} and \emph{O3N2} are given in function of the line ratios (i.e. 12+log(O/H) = p$_0$ + p$_1$ $\log_{10}$ R). In Table~\ref{tab:maiolino}, ~\ref{tab:jones} ~\ref{tab:curti} and ~\ref{tab:bian} we list all the coefficients for the different ratios used in this work.

\begin{table*}
\centering
\caption{\citealt{Maiolino2008} (M08) calibrations. The coefficients are listed in table 4 of that work and  were derived with data ranging between 7 and 9.2 in 12+log(O/H).}
\label{tab:maiolino}
\begin{tabular}{lllllll}
\hline
          & R                      & c$_0$   & c$_1$   & c$_2$   & c$_3$   & c$_4$   \\
\hline\hline
\emph{R23}       & (\oii+\oiiia+\oiiib)/\Hb & 0.7462  & -0.7149 & -0.9401 & -0.6154 & -0.2524 \\
\emph{O3}        & \oiiib/\Hb               & 0.1549  & -1.5031 & -0.9790 & -0.0297 & -       \\
\emph{O2}        & \oii/\Hb                 & 0.5603  & 0.0450  & -1.8017 & -1.8434 & -0.6549 \\
\emph{O32}       & \oii/\oiiib              & -0.2839 & -1.3881 & -0.3172 & -       & -       \\
\emph{O3N2}      & \oiiib / \nii            & 0.4520  & -2.6096 & -0.7170 & 0.1347  & -       \\
\emph{N2}        & \nii/\Ha                 & -0.7732 & 1.2357  & -0.2811 & -0.7201 & -0.3330 \\
\emph{Ne2O2}     & \neiii/\oii              & -1.2608 & -1.0861 & -0.1470 & -       & -     \\
\hline
\end{tabular}
\end{table*}

\begin{table*}
\centering
\caption{\citealt{Jones2015} (J15) calibrations, listed in table 1 of that work. The data used to derive this calibrations have metallicities from 7.6 to 9.0 in 12+log(O/H).}
\label{tab:jones}
\begin{tabular}{lllll}
\hline
          & R                        & c$_0$     & c$_1$   & c$_2$   \\
\hline\hline
\emph{R23}       & (\oii+\oiiia+\oiiib)/\Hb & -54.1003  & 13.9083 & -0.8782 \\
\emph{O3}        & \oiiib/\Hb               & -88.4378  & 22.7529 & -1.4501 \\
\emph{O2}        & \oii/\Hb                 & -154.9571 & 36.9128 & -2.1921 \\
\emph{O32}       & \oii/\oiiib              & 17.9828   & -2.1552 & -       \\
\emph{Ne2O2}     & \neiii/\oii              & 16.8974   & -2.1588 & -       \\
\emph{Ne2O3}     & \neiii/\oiiib            & -1.0854   & -0.0036 &  - \\     
\hline
\end{tabular}
\end{table*}

\begin{table*}
\centering
\caption{\citealt{Curti2017} (C17) calibrations. These coefficients are listed in table 2 of that work. The last column lists the range of applicability of each line ratio, as given by the authors. }
\label{tab:curti}
\begin{tabular}{llllllll}
\hline
         &R          & c$_0$  & c$_1$  & c$_2$  & c$_3$  & c$_4$  & Range                                    \\
\hline\hline
\emph{R23}       & (\oii+\oiiia+\oiiib)/\Hb & 0.527  & -1.569 & -1.652 & -0.421 & -      & 8.4 \textless 12+log(O/H) \textless 8.85 \\
\emph{O3}        & \oiiib/\Hb                                             & -0.277 & -3.549 & -3.593 & -0.981 & -      & 8.3 \textless 12+log(O/H) \textless 8.85 \\
\emph{O2}        & \oii/\Hb                                               & 0.418  & -0.961 & -3.505 & -1.949 & -      & 7.6 \textless 12+log(O/H) \textless 8.3  \\
\emph{O32}       & \oii/\oiiib                                            & -0.691 & -2.944 & -1.308 & -      & -      & 7.6 \textless 12+log(O/H) \textless 8.85 \\
\emph{O3N2}      & (\oiiib/\Hb) / (\nii/\Ha)                                         & 0.281  & -4.765 & -2.268 & -      & -      & 7.6 \textless 12+log(O/H) \textless 8.85 \\
\emph{N2}        & \nii/\Ha                                               & -0.489 & 1.513  & -2.554 & -5.293 & -2.867 & 7.6 \textless 12+log(O/H) \textless 8.85 \\
\hline
\end{tabular}
\end{table*}

\begin{table*}
\centering
\caption{\citealt{Bian2018} (BKD18) calibrations. The coefficients are listed in equations 11, 12 and 16 to 18 of that work. The first three calibrations are written in function of metallicity ($\log_{10} R = \sum_{n} c_n x^n$) and the last 3 in function of the line ratios (12+log(O/H) = p$_0$ + p$_1$ $\log_{10}$ R). The sample spans a metallicity range of 7.8 < in 12+log(O/H) < 8.4.  }
\label{tab:bian}
\begin{tabular}{llllll}
\hline
          & R                      & c$_0$   & c$_1$   & c$_2$  & c$_3$\\
\hline\hline
\emph{R23}       & (\oii+\oiiia+\oiiib)/\Hb & 138.0430 & -54.8284 &7.2954  & -0.32293  \\
\emph{O3}        & \oiiib/\Hb               &  43.9836 & -21.6211 & 3.4277 &  -0.1747 \\
\emph{Ne2O2}     & \neiii/\oii              &  7.80    & -0.63 & - & - \\
\hline
          & R                      & p$_0$   & p$_1$   &   & \\
\hline\hline
\emph{N2}        & \nii/\Ha   			    & 8.82 & 0.49& &  \\
\emph{O3N2}      & (\oiiib/\Hb) / (\nii/\Ha) &  8.97 & -0.39& & \\
\emph{O3}        & \oiiib/\Hb              &  8.54    & -0.59& &  \\
\hline
\hline
\end{tabular}
\end{table*}

\section{Line Fluxes}

The line fluxed presented here are based in the values available in the literature (see Table~\ref{tab:sample}). Using the literature flues, we have simultaneously derived extinction, electron temperature and density for each galaxy (see Table~\ref{tab:Te_and_ext}). We present the dust corrected fluxes in Table~\ref{tab:fluxes}, which were used to calculate electron temperature and density and the corrected oxygen fluxes in Table~\ref{tab:Te_and_ext}. Table~\ref{tab:ratios} shows the strong line ratios used to calculate metallicity using the strong line ratios method.

\begin{table*}
\caption{Extinction corrected line fluxes, normalised to \Hb.} 
\label{tab:fluxes}
\tabcolsep=0.25cm
\begin{tabular}{|lccccccc|}
\hline
Obj     &     	\oiiiUVa	& \oiiiUVb	& \ciiia	& \ciiib	& \neiii & \mbox{[O\,{\scshape ii]}}\,$\lambda$3727,29\,	& H$\delta$	\\ 
\hline\hline
CSWA20 	&0.062$\pm$0.538	&0.154$\pm$0.245	&0.193$\pm$0.136	&0.177$\pm$0.137	&0.421$\pm$0.136	&0.542$\pm$0.274	&-	\\ 
Abell\_860\_359 	&$\le$0.037	&0.101$\pm$0.273	&-	&-	&-	&$\le$1.265	&-	\\ 
Abell\_22.3 	&-	&-	&-	&-	&-	&1.490$\pm$0.194	&-	\\ 
RCSGA 	&-	&-	&-	&-	&0.308$\pm$0.004	&2.918$\pm$0.010	&0.265$\pm$0.004	\\ 
A1689\_31.1 	&0.185$\pm$0.002	&0.428$\pm$0.002	&0.302$\pm$0.002	&0.587$\pm$0.002	&0.199$\pm$0.003	&0.668$\pm$0.003	&-	\\ 
SMACS\_0304 	&0.026$\pm$0.140	&0.043$\pm$0.132	&-	&-	& 0.365$\pm$5.169	&2.613$\pm$0.326	&0.300$\pm$0.122	\\ 
MACS\_0451 	&0.243$\pm$0.099	&0.364$\pm$0.102	&-	&-	&-	& $\le$0.645	&-	\\ 
COSMOS\_12805 	&0.053$\pm$0.020	&0.081$\pm$0.028	&-	&-	&-	& $\le$2.924	&-	\\ 
BX660 	&$\le$0.000	&0.230$\pm$0.060	&-	&-	&-	&0.879$\pm$0.177	&-	\\ 
BX74 	&$\le$0.000	&0.169$\pm$0.175	&-	&-	&-	&1.125$\pm$0.313	&-	\\ 
BX418 	&$\le$0.000	&0.152$\pm$0.049	&-	&-	&-	&0.909$\pm$0.181	&-	\\ 
S16-stack 	&0.016$\pm$0.174	&0.046$\pm$0.120	&0.160$\pm$0.113	&0.111$\pm$0.117	&0.375$\pm$0.139	&2.677$\pm$0.326	&-	\\ 
COSMOS-1908 	&-	&-	&-	&-	&0.426$\pm$0.040	&0.509$\pm$0.045	&0.338$\pm$0.047	\\ 
Lynx arc 	& $\le$0.188	& $\le$0.396	& $\le$0.361	& $\le$0.248	& $\le$0.692	& $\le$0.251	&-	\\ 
SMACS\_2031 	&0.072$\pm$0.014	&0.218$\pm$0.018	&0.293$\pm$0.015	&0.205$\pm$0.022	&0.347$\pm$0.050	&0.703$\pm$0.071	&-	\\ 
SGAS\_1050 	&0.019$\pm$0.007	&0.054$\pm$0.007	&0.209$\pm$0.069	&0.126$\pm$0.079	&0.322$\pm$0.040	&0.795$\pm$0.021	&-	\\ 
\hline
\end{tabular}
\end{table*}

\begin{table*}
\contcaption{}
\tabcolsep=0.3cm
\begin{tabular}{|lccccccc|}
\hline
Obj     &     	 H$\gamma$	& \oiiiauroral	& H$\beta$	& \oiiia	& \oiiib	& H$\alpha$	& \nii	\\ 
\hline\hline
CSWA20 	&0.611$\pm$0.150	&0.053$\pm$0.356	&1.000$\pm$0.174	&1.651$\pm$0.237	&4.886$\pm$0.614	&2.869$\pm$0.371	&0.052$\pm$0.217	\\ 
Abell\_860\_359 	&-	&-	&1.000$\pm$0.096	&2.022$\pm$0.152	&6.083$\pm$0.420	&2.821$\pm$0.201	&-	\\ 
Abell\_22.3 	&-	&0.310$\pm$0.149	&1.000$\pm$0.142	&1.933$\pm$0.222	&6.225$\pm$0.634	&3.703$\pm$0.386	& $\le$0.037	\\ 
RCSGA 	&0.474$\pm$0.004	& $\le$0.052	&1.000$\pm$0.005	&1.489$\pm$0.006	&4.753$\pm$0.016	&2.720$\pm$0.010	&0.173$\pm$0.004	\\ 
A1689\_31.1 	&0.463$\pm$0.003	&0.168$\pm$0.003	&1.000$\pm$0.003	&1.438$\pm$0.004	&4.771$\pm$0.011	&-	&-	\\ 
SMACS\_0304 	&0.458$\pm$0.129	&-	&1.000$\pm$0.165	&1.315$\pm$0.192	&4.591$\pm$0.547	&2.868$\pm$0.352	&0.092$\pm$0.114	\\ 
MACS\_0451 	&-	&-	&1.000$\pm$0.136	&1.368$\pm$0.140	&3.943$\pm$0.381	&2.541$\pm$0.288	& $\le$0.064	\\ 
COSMOS\_12805 	&-	&-	&1.000$\pm$0.418	&1.889$\pm$0.694	&6.456$\pm$1.928	&2.824$\pm$0.836	& $\le$0.099	\\ 
BX660 	&-	&-	&1.000$\pm$0.277	&3.198$\pm$1.005	&6.395$\pm$1.289	&2.751$\pm$0.574	& $\le$0.070	\\ 
BX74 	&-	&-	&1.000$\pm$0.341	&2.449$\pm$0.726	&7.799$\pm$1.889	&3.097$\pm$0.760	& 0.107$\pm$0.210	\\ 
BX418 	&-	&-	&1.000$\pm$0.278	&2.299$\pm$0.822	&6.395$\pm$1.290	&2.792$\pm$0.582	& 0.149$\pm$0.042	\\ 
S16-stack 	&-	& $\le$0.066	&1.000$\pm$0.162	&1.434$\pm$0.200	&4.255$\pm$0.499	&2.866$\pm$0.346	&0.277$\pm$0.117	\\ 
COSMOS-1908 	&0.491$\pm$0.043	&0.124$\pm$0.049	&1.000$\pm$0.047	&2.270$\pm$0.081	&6.973$\pm$0.232	&-	&-	\\ 
Lynx arc 	&-	&-	& $\le$1.000	& $\le$2.579	& $\le$7.497	&-	&-	\\ 
SMACS\_2031 	&-	&-	&1.000$\pm$0.049	&1.446$\pm$0.054	&4.792$\pm$0.166	&-	&-	\\ 
SGAS\_1050 	&0.521$\pm$0.020	& $\le$0.014	&1.000$\pm$0.014	&2.349$\pm$0.030	&8.076$\pm$0.082	&-	&-	\\ 
\hline
\end{tabular}
\end{table*}

\begin{table*}
\caption{Strong line ratios used to calculate the metallicities in Section~\ref{sec:empirical} and \ref{sec:theo} (in log$_{10}$ scale). $\dagger$ Both \Hb\, and \oii\, are upper limits, so it is not possible to reliably calculate uncertainties.} 
\label{tab:ratios}
\tabcolsep=0.25cm
\renewcommand*{\arraystretch}{1.1}
\begin{tabular}{|lcccccccc|}
\hline
Obj     &     \emph{R23} &  \emph{N2} & \emph{O3} & \emph{O32} & \emph{O2} & \emph{O3\emph{N2}} & Ne3\emph{O3} & \emph{Ne3O2}\\
\hline\hline
CSWA20 &0.848$^{+0.063}_{-0.053}$ &-1.277$^{+0.313}_{-0.517}$ &0.686$^{+0.064}_{-0.055}$ &0.926$^{+0.232}_{-0.151}$ &-0.233$^{+0.161}_{-0.259}$ &1.526$^{+0.524}_{-0.317}$ &-0.121$^{+0.257}_{-0.133}$ &-1.055$^{+0.122}_{-0.219}$ \\ 
Abell\_860\_359 &0.971$^{+0.039}_{-0.033}$ &- &0.782$^{+0.039}_{-0.032}$ &0.683$^{+0.005}_{-0.006}$ &$\leq$0.098 &- &- &- \\ 
Abell\_22.3 &0.975$^{+0.053}_{-0.044}$ &$\geq$-2.003 &0.808$^{+0.049}_{-0.047}$ &0.776$^{+0.135}_{-0.107}$ &0.026$^{+0.121}_{-0.129}$ &2.111$^{+0.020}_{-0.023}$ &- &- \\ 
RCSGA &0.963$^{+0.001}_{-0.001}$ &-1.196$^{+0.008}_{-0.009}$ &0.676$^{+0.001}_{-0.001}$ &0.204$^{+0.001}_{-0.001}$ &0.471$^{+0.001}_{-0.001}$ &1.450$^{+0.008}_{-0.008}$ &-0.977$^{+0.004}_{-0.004}$ &-1.181$^{+0.005}_{-0.004}$ \\ 
A1689\_31.1 &0.837$^{+0.001}_{-0.001}$ &- &0.680$^{+0.001}_{-0.001}$ &0.871$^{+0.002}_{-0.002}$ &-0.190$^{+0.002}_{-0.002}$ &- &-0.526$^{+0.009}_{-0.007}$ &-1.397$^{+0.009}_{-0.007}$ \\ 
SMACS\_0304 &0.934$^{+0.049}_{-0.054}$ &-1.382$^{+0.263}_{-0.411}$ &0.666$^{+0.044}_{-0.056}$ &0.242$^{+0.024}_{-0.023}$ &0.420$^{+0.051}_{-0.056}$ &1.590$^{+0.416}_{-0.271}$ &0.137$^{+0.314}_{-0.505}$ &-0.089$^{+0.295}_{-0.502}$ \\ 
MACS\_0451 &0.774$^{+0.042}_{-0.042}$ &$\geq$-1.599 &0.596$^{+0.042}_{-0.042}$ &0.797$^{+0.006}_{-0.005}$ &$\geq$-0.201 & 1.783$^{+0.006}_{-0.005}$ &- &- \\ 
COSMOS\_12805 &1.053$^{+0.174}_{-0.116}$ &$\geq$-1.453 &0.809$^{+0.171}_{-0.109}$ &0.347$^{+0.020}_{-0.020}$ &$\leq$0.462 &1.809$^{+0.020}_{-0.020}$ &- &- \\ 
BX660 &1.028$^{+0.100}_{-0.095}$ &$\geq$-1.598 &0.811$^{+0.099}_{-0.079}$ &0.866$^{+0.030}_{-0.028}$ &-0.059$^{+0.101}_{-0.079}$ &1.960$^{+0.021}_{-0.021}$ &- &- \\ 
BX74 &1.048$^{+0.100}_{-0.065}$ &-1.211$^{+0.261}_{-0.572}$ &0.880$^{+0.098}_{-0.065}$ &0.801$^{+0.057}_{-0.051}$ &0.083$^{+0.102}_{-0.091}$ &1.636$^{+0.585}_{-0.244}$ &- &- \\ 
BX418 &0.972$^{+0.099}_{-0.087}$ &-1.085$^{+0.240}_{-0.511}$ &0.790$^{+0.108}_{-0.081}$ &0.846$^{+0.044}_{-0.045}$ &-0.050$^{+0.097}_{-0.097}$ &1.452$^{+0.494}_{-0.253}$ &- &- \\ 
S16-stack &0.920$^{+0.053}_{-0.043}$ &-1.045$^{+0.170}_{-0.223}$ &0.620$^{+0.055}_{-0.043}$ &0.183$^{+0.021}_{-0.021}$ &0.438$^{+0.054}_{-0.044}$ &1.230$^{+0.223}_{-0.164}$ &-0.863$^{+0.137}_{-0.161}$ &-1.048$^{+0.136}_{-0.159}$ \\ 
COSMOS-1908 &0.989$^{+0.016}_{-0.015}$ &- &0.844$^{+0.016}_{-0.015}$ &1.134$^{+0.036}_{-0.032}$ &-0.291$^{+0.037}_{-0.039}$ &- &-0.086$^{+0.057}_{-0.041}$ &-1.218$^{+0.038}_{-0.047}$ \\ 
Lynx arc &$\leq$1.014 &- &$\leq$0.875 &1.477$^{+0.017}_{-0.015}$ &-0.602$\dagger$ &- &0.461$^{+0.134}_{-0.241}$ &-1.020$^{+0.141}_{-0.244}$ \\ 
SMACS\_2031 &0.840$^{+0.018}_{-0.016}$ &- &0.680$^{+0.016}_{-0.015}$ &0.837$^{+0.042}_{-0.038}$ &-0.156$^{+0.043}_{-0.048}$ &- &-0.302$^{+0.073}_{-0.067}$ &-1.137$^{+0.052}_{-0.075}$ \\ 
SGAS\_1050 &1.050$^{+0.005}_{-0.004}$ &- &0.907$^{+0.005}_{-0.004}$ &1.009$^{+0.012}_{-0.010}$ &-0.102$^{+0.012}_{-0.013}$ &- &-0.388$^{+0.048}_{-0.059}$ &-1.398$^{+0.046}_{-0.062}$ \\
\hline
\end{tabular}
\end{table*}

\begin{table*}
\caption{Accuracy of the calibrations measured by the mean ($\overline{x}$) and standard deviation ($\sigma_{z\sim2}$) of the residuals 12+$\log$(O/H)$_{calib}$ - 12+$\log$(O/H)$_{Te}$, where the first is the metallicity obtained with that particular calibration and diagnostic and the second from the \Te method. The number of objects used to test each calibration are listed in the "\#" columns. The intrinsic dispersion ($\sigma_{int.}$) from \citealt{Jones2015} and \citealt{Curti2017} obtained at z$\sim0$ and up to z$\sim0.8$ respectively, is also listed for comparison. The last row lists the absolute mean of both $\overline{x}$ and $\sigma_{z\sim2}$ for each calibration.} 
\label{tab:stats}
\tabcolsep=0.14cm
\begin{tabular}{|lcccccccccccccc|}
\hline
     & \multicolumn{3}{c}{\citealt{Maiolino2008}} & \multicolumn{4}{c}{\citealt{Jones2015}} & \multicolumn{4}{c}{\citealt{Curti2017}} &\multicolumn{3}{c}{\citealt{Bian2018}} \\
& \# & $\overline{x}$ & $\sigma_{z\sim2}$ & \# &$\overline{x}$ & $\sigma_{z\sim2}$ & $\sigma_{int.}$ & \# & $\overline{x}$ & $\sigma_{z\sim2}$ & $\sigma_{int.}$ & \# & $\overline{x}$ & $\sigma_{z\sim2}$ \\
\hline\hline
\emph{R23} & 12 & -0.04$\pm$0.06 & 0.21$\pm$0.03 & 15 & 0.15$\pm$0.07 & 0.30$\pm$0.07 & 0.06 & 15 & 0.03$\pm$0.06 & 0.22$\pm$0.05 & 0.12 & 15 & 0.29$\pm$0.10 & 0.36$\pm$0.07\\
\emph{N2} & 10 & 0.13$\pm$0.07 & 0.23$\pm$0.04 & - & - & - & - & 10 & 0.15$\pm$0.07 & 0.21$\pm$0.05 & 0.10 & 10 & 0.16$\pm$0.06 & 0.19$\pm$0.05\\
\emph{O3} & 13 & -0.08$\pm$0.06 & 0.19$\pm$0.03 & 14 & 0.16$\pm$0.08 & 0.30$\pm$0.05 & 0.10 & 13 & 0.04$\pm$0.05 & 0.20$\pm$0.05 & 0.07 & 16 & 0.31$\pm$0.09 & 0.39$\pm$0.07\\
\emph{O32} & 16 & -0.24$\pm$0.11 & 0.41$\pm$0.06 & 16 & 0.05$\pm$0.07 & 0.27$\pm$0.05 & 0.23 & 16 & -0.01$\pm$0.08 & 0.33$\pm$0.06 & 0.14 & 16 & 0.15$\pm$0.07 & 0.28$\pm$0.05\\
\emph{O2} & 16 & 0.23$\pm$0.06 & 0.24$\pm$0.04 & 15 & 0.06$\pm$0.06 & 0.25$\pm$0.04 & 0.15 & 16 & -0.01$\pm$0.06 & 0.25$\pm$0.05 & 0.26 & - & - & -\\
\emph{O3N2} & 10 & 0.15$\pm$0.07 & 0.21$\pm$0.05 & - & - & - & - & 10 & 0.19$\pm$0.06 & 0.21$\pm$0.06 & 0.09 & 10 & 0.16$\pm$0.06 & 0.21$\pm$0.06\\
\emph{Ne3O2} & 9 & -0.15$\pm$0.17 & 0.48$\pm$0.06 & 9 & 0.02$\pm$0.09 & 0.27$\pm$0.06 & 0.22 & - & - & -& - & 9 & -0.01$\pm$0.11 & 0.36$\pm$0.07 \\
mean & & 0.15 & 0.28 & & 0.09 & 0.28 & & & 0.07 & 0.24 & & & 0.18 & 0.30 \\
\hline
\end{tabular}
\end{table*}


\bsp	
\label{lastpage}
\end{document}